\documentclass[prb,twocolumn,showpacs,preprintnumbers,amsmath,aps]{revtex4}
\usepackage{graphicx,hyperref}% Include figure files
\usepackage{xcolor}
\usepackage[normalem]{ulem}
\usepackage{color}
\usepackage{bm}% bold math
\usepackage{subfigure}% bold math

\def\nbC{{\mathchoice {\setbox0=\hbox{$\displaystyle\rm C$}%
\hbox{\hbox to0pt{\kern0.4\wd0\vrule height0.9\ht0\hss}\box0}}
{\setbox0=\hbox{$\textstyle\rm C$}\hbox{\hbox
to0pt{\kern0.4\wd0\vrule height0.9\ht0\hss}\box0}}
{\setbox0=\hbox{$\scriptstyle\rm C$}\hbox{\hbox
to0pt{\kern0.4\wd0\vrule height0.9\ht0\hss}\box0}}
{\setbox0=\hbox{$\scriptscriptstyle\rm C$}\hbox{\hbox
to0pt{\kern0.4\wd0\vrule height0.9\ht0\hss}\box0}}}}
\def\nbQ{{\mathchoice {\setbox0=\hbox{$\displaystyle\rm
Q$}\hbox{\raise
0.15\ht0\hbox to0pt{\kern0.4\wd0\vrule height0.8\ht0\hss}\box0}}
{\setbox0=\hbox{$\textstyle\rm Q$}\hbox{\raise
0.15\ht0\hbox to0pt{\kern0.4\wd0\vrule height0.8\ht0\hss}\box0}}
{\setbox0=\hbox{$\scriptstyle\rm Q$}\hbox{\raise
0.15\ht0\hbox to0pt{\kern0.4\wd0\vrule height0.7\ht0\hss}\box0}}
{\setbox0=\hbox{$\scriptscriptstyle\rm Q$}\hbox{\raise
0.15\ht0\hbox to0pt{\kern0.4\wd0\vrule height0.7\ht0\hss}\box0}}}}
\def\nbT{{\mathchoice {\setbox0=\hbox{$\displaystyle\rm
T$}\hbox{\hbox to0pt{\kern0.3\wd0\vrule height0.9\ht0\hss}\box0}}
{\setbox0=\hbox{$\textstyle\rm T$}\hbox{\hbox
to0pt{\kern0.3\wd0\vrule height0.9\ht0\hss}\box0}}
{\setbox0=\hbox{$\scriptstyle\rm T$}\hbox{\hbox
to0pt{\kern0.3\wd0\vrule height0.9\ht0\hss}\box0}}
{\setbox0=\hbox{$\scriptscriptstyle\rm T$}\hbox{\hbox
to0pt{\kern0.3\wd0\vrule height0.9\ht0\hss}\box0}}}}
\def\nbS{{\mathchoice
{\setbox0=\hbox{$\displaystyle     \rm S$}\hbox{\raise0.5\ht0%
\hbox to0pt{\kern0.35\wd0\vrule height0.45\ht0\hss}\hbox
to0pt{\kern0.55\wd0\vrule height0.5\ht0\hss}\box0}}
{\setbox0=\hbox{$\textstyle        \rm S$}\hbox{\raise0.5\ht0%
\hbox to0pt{\kern0.35\wd0\vrule height0.45\ht0\hss}\hbox
to0pt{\kern0.55\wd0\vrule height0.5\ht0\hss}\box0}}
{\setbox0=\hbox{$\scriptstyle      \rm S$}\hbox{\raise0.5\ht0%
\hboxto0pt{\kern0.35\wd0\vrule height0.45\ht0\hss}\raise0.05\ht0%
\hbox to0pt{\kern0.5\wd0\vrule height0.45\ht0\hss}\box0}}
{\setbox0=\hbox{$\scriptscriptstyle\rm S$}\hbox{\raise0.5\ht0%
\hboxto0pt{\kern0.4\wd0\vrule height0.45\ht0\hss}\raise0.05\ht0%
\hbox to0pt{\kern0.55\wd0\vrule height0.45\ht0\hss}\box0}}}}
\def\nbZ{{\mathchoice {\hbox{$\sf\textstyle Z\kern-0.4em Z$}}
{\hbox{$\sf\textstyle Z\kern-0.4em Z$}}
{\hbox{$\sf\scriptstyle Z\kern-0.3em Z$}}
{\hbox{$\sf\scriptscriptstyle Z\kern-0.2em Z$}}}}
\begin{document}

\title{On the anomalous elasticity in the mechanical response of amorphous solids}

\author{Gilles Tarjus} \email{tarjus@lptmc.jussieu.fr}
\affiliation{LPTMC, CNRS-UMR 7600, Sorbonne Universit\'e,
4 Pl. Jussieu, 75252 Paris cedex 05, France}

\author{Misaki Ozawa} \email{misaki.ozawa@univ-grenoble-alpes.fr}
\affiliation{Univ. Grenoble Alpes, CNRS, LIPhy, 38000 Grenoble, France}

\author{Giulio Biroli} \email{giulio.biroli@lps.ens.fr}
\affiliation{Laboratoire de Physique de l'Ecole Normale Sup\'erieure, ENS, Universit\'e PSL, CNRS, Sorbonne Universit\'e, Universit\'e Paris Cit\'e, F-75005 Paris, France}

\date{\today}

\begin{abstract}
The response of amorphous solids to a mechanical perturbation consists in an elastic and a plastic deformation. The latter is mediated by localized irreversible rearrangements associated with  Eshelby-like quadrupolar singularities in the displacement field. It has recently been argued that a density of such singularities leads to an anomalous elastic behavior taking the form of screening effects, which goes beyond classical elastic predictions. Here, we reexamine this scenario using general theoretical arguments and a description in terms of an  elasto-plastic model, which we compare with atomistic simulations of the canonical Eshelby inclusion geometry. We discuss the conditions under which a finite, {\it i.e.}, nonvanishing, density of quadrupolar events is created by an imposed perturbation. We argue that, except when the perturbation is macroscopic, there are many situations in which the density of quadrupolar defects is zero in the thermodynamic limit. In these cases, we find that plastically active quadrupoles emerge in a region whose size generically scales as the spatial extent $\ell$ of the mechanical perturbation. This mechanism leads to anomalous elasticity on a scale $\ell$ close to the perturbation and to conventional elasticity beyond. The simulations of the elasto-plastic model reproduce the emergence of plastic quadrupoles in a region set by $\ell$ and the associated renormalization of the effective shear modulus, but they do not exhibit the dipole-screening signatures reported in atomistic and experimental studies. Our analysis delineates the scale-dependent breakdown of long-wavelength elasticity in amorphous materials and suggests directions for incorporating anomalous screening into mesoscopic modeling frameworks.
\end{abstract}

\pacs{11.10.Hi, 75.40.Cx}

\maketitle

\section{Introduction}

Contrary to crystalline solids whose response to deformations can be purely elastic, amorphous solids, which comprise a very large class of materials ranging from dense granular suspensions and soft-condensed matter systems to metallic and silicate glasses, always respond through a combination of elastic (reversible) and plastic (irreversible) behavior.\cite{argon79,falk98,malandro99,maloney04,EPM_review} 
Localized plastic events, which take the form of quadrupolar singularities in the displacement field akin to those generated by an Eshelby inclusion,\cite{eshelby57} appear in a macroscopic amorphous material for any infinitesimal deformation\cite{karmakar10,hentschel11} and 
have a drastic influence on its nonlinear mechanical response. 

Recently, Procaccia and coworkers introduced the notion of ``anomalous elasticity'' to describe a number of unusual features of the quasi-linear response of amorphous solids that are not described by the classical theory of elasticity.\cite{lemaitre_anomalous,anomalous_review} 
It is argued that the presence of Eshelby-like quadrupolar singularities associated with plastic events induces screening effects in the response of the solid. If the distribution of such quadrupoles has a low density and is spread rather uniformly in space, screening simply leads to a renormalization of the linear elastic constants. For a higher density and a nonuniform distribution of the quadrupoles, dipoles corresponding to gradients of the quadrupolar field lead to more striking screening effects that are at odds with conventional elasticity theory. On the theoretical side, the approach is based on energy minimization starting from a quadratic Lagrangian description of the elastic response in the presence of a quadrupolar field associated with the plastic events.\cite{hentschel_energy,lemaitre_anomalous,anomalous_3D} The predictions appear supported by both atomistic numerical simulations and experiments on granular materials.\cite{anomalous_expt,anomalous_2Dglass,anomalous_dipole-meas,anomalous_dipole-shear,anomalous_odd,anomalous_flattening}

The goal of this paper is to provide a complementary study of anomalous elasticity. We proceed in two ways. First, on a general theoretical ground, we discuss when and where a nonzero density of quadrupolar Eshelby-like singularities is actually present in an amorphous solid under loading and can then lead to an anomalous elastic response to mechanical perturbations. Our aim is to address the basic tenets of anomalous elasticity and to understand the role played by the spatial extent of the mechanical perturbations and by the system size. This part tries to clarify to what extent the long-wavelength elasticity theory is modified and becomes anomalous.  
Our second aim is to study the emergence of anomalous elasticity within elasto-plastic models. The latter represent a coarse-grained mesoscopic description of the deformation of amorphous solids.\cite{EPM_review} We focus on the limit of very low temperature and very slow deformation corresponding to the athermal quasi-static (AQS) limit of the deformation protocols. In this limit, elasto-plastic models are cellular automata defined by kinetic rules. The prime interest of such models is to account for the dynamical evolution of the system (under deformation) among a potentially very long sequence of mechanical equilibria. The down side is a rather crude description of what actually happens in an elementary step when the system moves from one specific equilibrium to the next. We would therefore like to assess if the anomalous elasticity that has been discovered in such elementary steps is at least qualitatively included in elasto-plastic models and, if not, what could be the options for an improved description. We will investigate this issue in the case of the so-called Eshelby problem in a simple 2-dimensional elasto-plastic model and compare the outcome with results of atomic simulations and theoretical considerations.\cite{hentschel_eshelby,avanish-eshelby} 
\\

The paper is organized as follows. In Sec.~\ref{sec_finite-size} we stress that anomalous elasticity requires the existence of a nonzero density of quadrupolar Eshelby-like singularities and we discuss the conditions under which this can occur. In Sec.~\ref{sec_anomalous-EPM} we provide a comparative presentation of the formalism used by Procaccia and coworkers to describe anomalous elasticity in amorphous solids and of elasto-plastic models based on kinetic rules. We assess the ability of such models to capture phenomena associated with anomalous elasticity in the example of the Eshelby problem in Sec.~\ref{sec_Eshelby}. Finally, we give some concluding remarks  in Sec.~\ref{sec_conclusion} and we provide additional details on the elasto-plastic model used in this work in an Appendix.
\\

\section{Anomalous elasticity: Is there a breakdown of classical elasticity on all length scales?}
\label{sec_finite-size}

Classical elasticity is a long-wavelength theory: it describes mechanical responses on length-scales well above the microscopic scale all the way up to the system size.  Anomalous elasticity signals a breakdown of this continuum description. Here, we wish to understand the spatial extent of this breakdown.\cite{footnote_anomalous} In an infinite system, does a perturbation introduced on a mesoscopic scale create distortions that never recover the standard elastic form, no matter how far away we look?  Or is conventional elasticity eventually restored at large enough distances?  In simulations and experiments, the sample is finite, and its size may be comparable to that of the imposed perturbation.  What is then the interplay between these two length scales?  The aim of this section is to address and answer these questions.

What is specific about the mechanical response of amorphous solids is the ubiquitous presence of swift local irreversible rearrangements of the constituents that correspond to elementary plastic events. These events are localized and can be associated with quadrupolar Eshelby-like singularities in the displacement field. Anomalous elasticity stems from these quadrupolar singularities which are always present in a macroscopic amorphous solid under load.\cite{lemaitre_anomalous} However, the point we would like to make is that for anomalous elasticity to be observable in a material on macroscopic scales, whether in the simple form of a renormalization of the elastic constants or for more exotic dipole screening effects, there should be {\it a nonzero density of localized plastic singularities} or, equivalently, an extensive number of them. 
This requirement is not necessarily fulfilled. As we explain below, unless the mechanical perturbation is also macroscopic, there is indeed a breakdown of classical elasticity but not in the bulk of the sample, away from the perturbation. We try to make this clearer and refine the argument in what follows.

The fact that a nonzero density of {\it localized} defects is needed to have observable consequences in a macroscopic sample can trivially be illustrated in the context of systems used as an analogy by Procaccia, Moshe, and their coworkers.\cite{lemaitre_anomalous,anomalous_review,anomalous_3D,anomalous_interm-phase,moshe_metamaterials,moshe23} Consider, for instance, the 2-dimensional XY model where the physics of the BKT transition involves vortices which are topological defects in the ferromagnetic order (or equivalently, melting of 2-dimensional solids where dislocations in the hexagonal order are akin to vortices)\cite{nelson_2Dreview}. Bound vortices do induce a renormalization of the stiffness constant that characterizes the low-temperature quasi-ordered phase, but this requires that vortices are present with a nonzero density in the system. Similarly, in electrostatics,\cite{jackson_electrostatics} the renormalization of the dielectric constant in a material subject to an applied electric field requires a nonzero density of induced local dipoles, as produced, for instance, by an extensive number of polarizable molecules.

In the physical situations considered by Procaccia {\it et al.}, one perturbs the solid from a mechanical equilibrium, let it relax to a new mechanical equilibrium, and consider the affine and nonaffine components of the displacement field between the two energy minima (in an experiment at zero temperature). When and where can a nonzero density of localized (quadrupolar) plastic events appear under such 
conditions? 

Plastic events are generated both directly by the elastic response of the material that increases the stress(es) and may subsequently induce stress-releasing local reorganizations and by the occurrence of other plastic events that may trigger new reorganizations due to the ensuing long-distance deformation: see, {\it e.g.}, [\onlinecite{EPM_review}]. Our contention is that, {\it generically}, neither of the two mechanisms, nor their combination, can produce a nonzero density of localized quadrupolar-like plastic events in the bulk of the material.

There are, however, cases in which this occurs. It is useful to discuss them to clarify what is needed to have a finite density of quadrupolar-like plastic events:
\begin{itemize}
\item {\it Jamming.} A first potential example is in the jammed phase of, say, a frictionless assembly of spherical particles at a vanishing distance of the jamming transition. There, the system is isostatic and marginal and an extensive number of plastic events could be triggered by a small perturbation at the transition. Nonetheless, as shown by Fu {\it et al.},\cite{anomalous_finite-size} the intermediate phase that is found in a finite-size system between the unjammed system and the quasi-elastic solid and is associated with an anomalous dipole screening behavior\cite{anomalous_interm-phase,anomalous_interm-phase3D} vanishes in the thermodynamic limit (when keeping the scale of the perturbation fixed). This can be understood via a simple scaling argument in which one compares the isostaticity length, which diverges when approaching the jamming point, and the dipole screening length, which scales as the system size.\cite{anomalous_interm-phase,anomalous_finite-size} (Note that because it is proportional to the system size this screening length does not really introduce a new scale in the problem. The phenomenon of dipole screening has also recently been challenged from a different theoretical perspective in [\onlinecite{bulbul25}]. )

\item {\it Yielding of brittle materials.} A different example is provided by a brittle material under uniform shear when it is precisely at its yielding transition which in AQS 
conditions leads to a jump of order 1 in the average stress.\cite{ozawa14} At such a yielding transition the energy minimization for an infinitesimal strain increase involves the development of a shear band accompanied by an extensive number of quadrupolar Eshelby-like plastic events.\cite{ozawa14,procaccia_shearband,ozawa22} 

\item {\it Polarizability under strain.} Another case, not yet observed in simulations or experiments, involves an additional mechanism of polarizability under strain. An extensive 
number of quadrupolar defects has been predicted in assemblies of polarizable quadrupolar Eshelby-like singularities under strain beyond a critical pressure.\cite{hentschel_shear-polarizable}

\item {\it Macroscopic perturbations.} A final and more important situation is when the deformation is macroscopic, {\it i.e.} applied over the scale of the whole system, and is of order 1. This is, {\it e.g.}, the case when a spherical cap made of an amorphous solid is suddenly flattened\cite{anomalous_flattening} or in a Couette geometry\cite{procaccia_couette} when the inner and the outer rotating cylinders are both of macroscopic extent. 
\end{itemize}

All of the above counterexamples involve either a perturbation on the scale of the system (and of magnitude of order 1) or a transition of some sort triggering a collective reaction in the medium that may generate a nonzero density of plastic events. However, in many situations, the density of quadrupolar plastic defects induced in the bulk of the amorphous solid is zero in the thermodynamic limit. In such instances, anomalous elasticity is then a finite-size effect, but it is worth discussing in detail how this comes about and with which consequences. This will provide a microscopic justification for  the above statements. 

We will focus on three situations studied by Procaccia and coworkers: the response to the inflation of a central disk or sphere in a spherical geometry,\cite{lemaitre_anomalous,anomalous_3D,anomalous_expt,anomalous_2Dglass,anomalous_dipole-meas,anomalous_interm-phase,anomalous_finite-size} the Eshelby problem that consists in the distortion of a central disk to an ellipse,\cite{hentschel_eshelby} and the response to a uniform simple shear.\cite{anomalous_dipole-shear} For all cases, we call $L$ the linear system size (which in the spherical geometry is the outer radius $r_{\rm out}$ in which the solid is contained) and we characterize the scale of the distortion by $\ell$. In the uniform-shear setting, this is the distance over which one boundary plate is displaced with respect to the other and the corresponding applied strain is $\ell/L$. In the inflation experiment and in the Eshelby problem, $\ell^d$, where $d$ denotes the number of spatial dimensions, is associated with the volume of the central object, $r_{\rm in}^d$, multiplied by either the relative radius change $d_0/r_{\rm in}$ (inflation) or the relative distortion $\delta$ (Eshelby problem).

Consider first how the elastic response can trigger plastic events. In the uniform-shear setting, the stress increase in any small local region of the system (assuming some coarse-graining) scales as $\ell/L$ while in the other cases, assuming an elastic $r^{-d}$ stress propagator and overlooking all angular and tensorial effects that could only reduce the influence of the perturbation, it scales as $\ell^d/r^d$ times the relative amplitude which we take of order 1 (for deriving an upper bound). If the excitations of the system, which take here the form of localized plastic rearrangements, are gapped, no plastic events are generated unless $\ell$ scales as the system size $L$. If the excitations have a pseudo-gap characterized by an exponent $\theta$,\cite{wyart-muller} the probability that a local region yields plastically goes as $(\ell/L)^{\theta+1}$ at each site for a uniform shear and as $(\ell/r)^{d(\theta+1)}$ for all sites at a (large enough) distance $r$ from the central perturbation in the inflation and Eshelby problems. As a result, the density of plastic defects directly created through the elastic response in the bulk of the material goes either as $(\ell/L)^{\theta+1}$ or $(\ell/L)^{d(\theta+1)}$ and therefore vanishes in the thermodynamic limit $L\to \infty$ except if $\ell$ scales as the system size.

Of course, the response also involves the effect of plastic events  already activated. This is known to trigger a collective phenomenon in the form of a system-spanning avalanche of plastic events. However, such avalanches are generically {\it subextensive} (except at a brittle yielding transition or at the jamming critical point: see above), so that the density of plastic events in the whole material still goes to zero in the thermodynamic limit: see,  {\it e.g.}, [\onlinecite{lerner09_avalanches}].

The conclusion of the above discussion is that, barring exceptional situations such as the jamming transition of a granular medium or the brittle yielding transition of a sheared amorphous solid, a nonzero density of quadrupolar Eshelby-like singularities (plastic events) leading to anomalous elasticity in the bulk of the material can only take place when the scale of the perturbation is proportional to the system size. If this is not the case, classic elastic behavior is recovered far enough from the perturbation. 

Does this still imply a breakdown of classical elasticity as a long wave-length theory? Actually, yes.
Classical elasticity is intended to be a universal, coarse-grained (or renormalized) description. Once lengths below the microscopic cutoff $a$ are integrated out, all microscopic details disappear and only material-specific elastic constants remain, just as transport coefficients remain in hydrodynamics. As a result, classical elastic behavior should hold on any scale much larger than $a$. In the examples discussed above, this is not what happens.   In the Eshelby problem, for instance, a perturbation of size $\ell\gg a$ triggers a finite density of quadrupolar rearrangements and produces anomalous behavior over distances of order $\ell$.  Ordinary elasticity should apply even near the inclusion, provided that we probe scales much larger than the microscopic length $a$. Instead, the distance at which standard behavior is recovered grows with $\ell$ itself. Therefore, classical elasticity falls short of providing a proper large-scale theory for amorphous solids. The most spectacular failure is that it does not correctly predict the macroscopic response to a macroscopic load, \cite{anomalous_flattening} the very regime for which it was originally devised.

\section{Anomalous elasticity and elasto-plastic models}
\label{sec_anomalous-EPM}

\subsection{Recap of the formalism of Procaccia {\it et al.}}
\label{sub_recap}

To assess the ability of elasto-plastic models to capture (or not) the phenomenon of anomalous elasticity in amorphous solids under load, we first recall the formalism used to derive the screening phenomena that characterize this anomalous elastic response.\cite{hentschel_energy,lemaitre_anomalous,anomalous_3D} As before, we consider an AQS protocol. The system is perturbed from one energy minimum and then relaxes to a new minimum. This process is described via the minimization of a quadratic energy functional, 
which therefore corresponds to a linear or quasi-linear mechanical response. The response has a purely elastic component described in the continuum limit by a bare displacement field, and the associated elastic strain, $\bf u^{\rm el}(\bf r)$, and stress, $\bm \sigma^{\rm el}(\bf r)$, fields, as well as a plastic response heuristically treated through the introduction of a quadrupole field $\bf Q(\bf r)$. The latter corresponds to the Eshelby-like quadrupolar singularities created in the displacement field when the system relaxes to the new mechanical equilibrium. The energy to be minimized reads\cite{lemaitre_anomalous,hentschel_energy}
\begin{equation}
\begin{aligned}
\label{eq_lagrangian}
\mathcal U[\bf u^{\rm el},\bf Q]=&\frac 12 \int_{\bf r} \bm\sigma^{\rm el}(\bf r) {\bf :} {\bf u}^{\rm el}(\bf r) + \int_{\bf r} {\bf u}^{\rm el}(\bf r) {\bf :} 
\bm \Gamma {\bf :} {\bf Q}(\bf r)  \\&
+ \frac 12 \int_{\bf r \bf r'}  {\bf Q}(\bf r) {\bf :} \bm \Lambda(\bf r-\bf r') {\bf:} {\bf Q}(\bf r') ,
\end{aligned}
\end{equation}
where $\int_{\bf r}\equiv \int d^d\bf r$, $\bm \Gamma$ is a  phenomenological (constant) tensor characterizing the coupling between elastic strain and plastic events, $\bm \Lambda(\bf r-\bf r')$ is the elastic propagator describing the long-range interaction between two Eshelby-like quadrupoles, and the elastic stress is linearly related to the elastic strain as
\begin{equation}
\label{eq_stress-strain_elastic}
\bm\sigma^{\rm el}(\bf r)={\bf A} {\bf :} {\bf u}^{\rm el}(\bf r)
\end{equation}
with $\bf A$ the bare (constant) elastic tensor. From the quadrupole field, one can also define a plastic strain and a plastic stress\cite{lemaitre_anomalous}
\begin{equation}
\begin{aligned}
\label{eq_stress+strain_plastic}
&{\bf u}^{\rm pl}(\bf r)= \int_{\bf r'}  {\bf G}^{(u)}(\bf r-\bf r') {\bf :} {\bf Q}(\bf r'),\\&
\bm\sigma^{\rm pl}(\bf r)=\int_{\bf r'}  {\bf G}^{(\sigma)}(\bf r-\bf r') {\bf :} {\bf Q}(\bf r') ,
\end{aligned}
\end{equation}
which can be combined with the (bare) elastic components to give the full strain and stress tensor fields. In the above expressions, ${\bf G}^{({\bf u})}$ and ${\bf G}^{(\sigma)}$ are the strain and stress elastic Green's functions, respectively.

Note that (i) all quantities {\it a priori} have a tensorial character, (ii) the minimization is not with respect to the elastic strain but to the displacement field (on top of the quadrupole field), and (iii) the elastic propagator $\bm \Lambda(\bf r-\bf r')$  diverges as $\vert {\bf r}- {\bf r'} \vert \to 0$. The latter is an artifact of the continuum limit and is simply cured by introducing a UV cutoff $a$, which physically represents the core size of the quadrupoles, and by replacing the expression of the propagator by $\bm \Lambda_0\delta^{(d)}({\bf r}-{\bf r'}) +\theta(\vert {\bf r}-{\bf r'}\vert-a)\bm \Lambda(\bf r-\bf r')$ where $\theta$ is the Heaviside step function and $\bm \Lambda_0$ a phenomenological parameter associated with some core energy for the quadrupoles.

When the density of quadrupoles is small (but nonzero: see above) and uniform, it leads to a renormalization of the elastic tensor $\bf A$ in an otherwise unchanged elastic description.\cite{footnote_literature} Yet, when the density is higher and the spatially nonuniform distribution of the quadrupoles cannot be neglected, the lowest-order effect is due to the gradient of the quadrupole field (its divergence), which represents a dipole field. Anomalous screening effects not accounted for by the conventional elasticity theory are then predicted to appear. In both cases, the effect is quadratic in the coupling tensor $\bm \Gamma$.\cite{lemaitre_anomalous,anomalous_3D}

\subsection{Elasto-plastic models}
\label{sub_EPM}

Elasto-plastic models provide a description that is quite different than the minimization of a quadratic energy functional just discussed. They focus on the full evolution of the deformed amorphous solid which, in an AQS protocol, implies a whole trajectory among energy minima, not just the process of relaxing to a specific one after a sudden perturbation. However, this comes at the cost of simplifying assumptions. A review is provided in [\onlinecite{EPM_review}].

Elasto-plastic models describe the amorphous solid as divided into mesoscopic blocks whose size is roughly the typical extent of a plastic rearrangement. Each block $i$ centered on ${\bf r}_i$ is characterized by two dynamical variables, a local stress which, for simplicity, we will take as a scalar $\sigma_i$, representing the most relevant component of the stress tensor, {\it e.g.}, the XY component in a simple shear protocol along the $x$-direction, and the plastic activity indicator $n_i$ which is equal to $0$ if the block has no irreversible rearrangement and to $1$ otherwise. In the AQS driving, the model is a cellular automaton characterized by kinetic rules that describe the evolution of the local variables $\sigma_i$ and $n_i$ through a combination of elastic and plastic responses. We illustrate the rules for the application of a simple shear strain increment 
$\delta\gamma$:\cite{EPM_review}

- A plastically inactive block ($n_i=0$) displays a purely elastic response to the applied deformation, leading to
\begin{equation}
\delta\sigma^{\rm el/appl}_i=\mu \delta\gamma,
\end{equation}
where $\mu$ is the shear modulus (for a homogeneous and isotropic medium), and a stress change due to the elastic propagation of plastic events occurring elsewhere in the material,
\begin{equation}
\delta\sigma^{\rm el/pl}_i=\mu \sum_{j\neq i}G_{ij}^{\rm Eshelby}\Delta\sigma_j n_j,
\end{equation}
where $\Delta\sigma_j$ is the stress drop characterizing the local yielding event of the plastically active block $j$ and $G_{ij}^{\rm Eshelby}$ is the long-ranged Eshelby stress propagator describing the redistribution of stress from site $j$ to site $i$; in 2 dimensions for instance, this propagator is given by $\cos(4\theta_{ij})/(\pi r_{ij}^2)$ where $\theta_{ij}$ is the angle between the separation vector ${\bf r}_{ij}={\bf r}_i-{\bf r}_j$ and the $x$-axis (the direction of the deformation) and $r_{ij}=|{\bf r}_{ij}|$ is the distance between the two sites.

- An active block ($n_i=1$) yields and releases its stress,
\begin{equation}
\delta\sigma_i=-\Delta\sigma_i, 
\end{equation}
where $\Delta\sigma_i$ is drawn from a given distribution. In the AQS setting, the recovery ($n_i=1 \to n_i=0$) is instantaneous.

- A block switches from inactive to active ($n_i=0 \to n_i=1$) when its local stress becomes larger than a fixed threshold $\sigma^{\rm th}$, which, for simplicity, we take equal for all blocks.

These rules are supplemented by initial conditions taken from a distribution $P_0(\{\sigma_i\})$ that characterizes the state of the solid prior to deformation.  As already stressed, in the simplified description given here, the tensorial character of the fields is overlooked, but more involved tensorial elasto-plastic models can be implemented if needed.
\\

As such, an elasto-plastic model has no displacement field, only the stress field being described (on a lattice instead of in the continuum, but this is not crucial), and no energy minimization (although the system is assumed to evolve from one mechanical equilibrium to another). The main ingredients that are discussed in the previous subsection and lead to anomalous elasticity are nevertheless present: the combination of elastic and plastic responses, the quadrupole field that corresponds to the active sites $n_i=1$ with the quadrupole amplitude given by the stress drop $\Delta\sigma_i$, {\it i.e.},
\begin{equation}
\label{eq_EPM_quadrupole} 
Q({\bf r})\equiv \sum_i \delta^{(d)}({\bf r}-{\bf r}_i) \Delta\sigma_i n_i,
\end{equation}
the long-range effect of plastic events on the whole material, and their interaction in the sense that a plastic event can trigger other plastic events. In addition, the quadrupole field represented by the active sites can be nonuniform and therefore lead to dipoles in the form of (discrete) gradients.  

In elasto-plastic models, and this is what makes them a useful tool for modeling deformation of amorphous solids, plastic events are generated and die, and their configuration evolves with deformation. Furthermore, their density may be zero or nonzero in the thermodynamic limit. Assessing whether elasto-plastic models also capture some aspects of the anomalous elasticity (when present) is therefore important. In the following, we present an investigation for the case of the Eshelby problem already introduced.
\\

\begin{figure}
\includegraphics[width=0.45\linewidth]{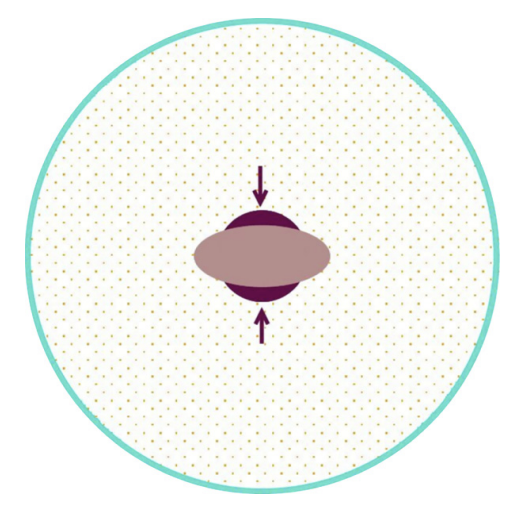}
\includegraphics[width=0.4\linewidth]{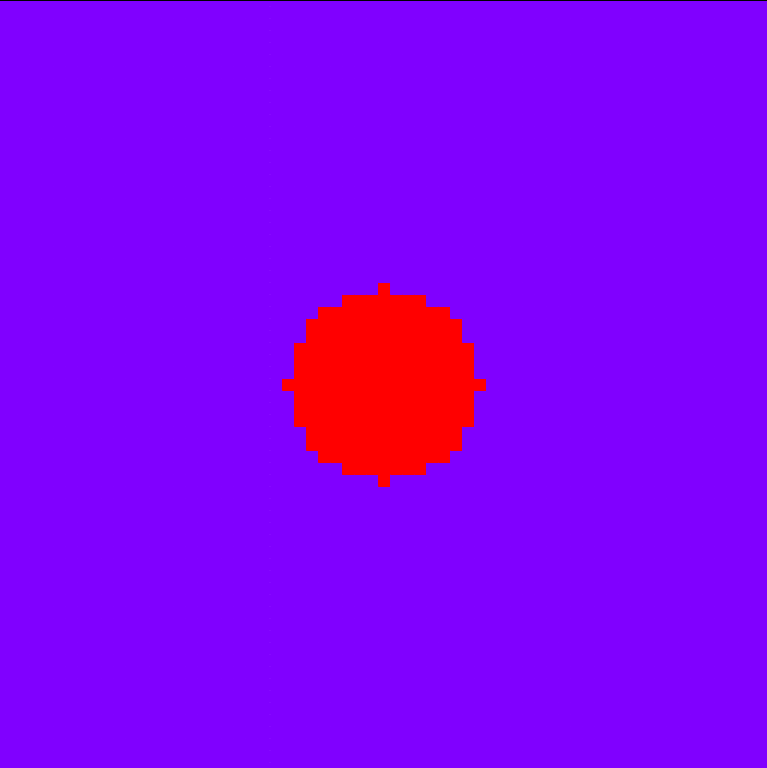}
\caption{Sketch of the Eshelby problem: (a) In the atomic simulations of [\onlinecite{hentschel_eshelby}] a disk of radius $r_{\rm in}=5$ is instantaneously deformed into an ellipse of the same area with a deformation parameter $\delta=1.05$ in a sample bounded by an outer circle 
of radius $r_{\rm out}=80$ (the unit of length is the average atomic diameter).
(b) In the present elasto-plastic model description, a large central plastic defects of diameter $\ell$ on the verge of yielding mimics the effect of an Eshelby inclusion in a system of linear size $L$ (with periodic boundary conditions).
}
\label{fig_sketch-Eshelby}
\end{figure}

\section{Illustration: The Eshelby problem}
\label{sec_Eshelby}

\subsection{The setup in atomic simulation and elasto-plastic model}
\label{sub_setup}

The Eshelby problem in 2 dimensions was investigated by atomic computer simulation and theoretical arguments in [\onlinecite{hentschel_eshelby}]. 
There, an inner disk of radius $r_{\rm in}$ at the center of an amorphous solid sample was instantaneously deformed into an ellipse of the same area with a relative deformation parameter $\delta$ and the subsequent response of the solid, which is contained in a domain bounded by an outer circle of radius $r_{\rm out}$, was studied: see the sketch in Fig.~\ref{fig_sketch-Eshelby}(a). The displacement field induced by the deformation was predicted by the theory discussed in Sec.~\ref{sub_recap} and was successfully compared with simulation results for glass samples quenched from different parent temperatures. The better annealed system (lowest parent temperature) displays anomalous elasticity in the form of a renormalization of the elastic constant, while the poorly annealed one (highest parent temperature) further shows anomalous dipole screening. Note that $r_{\rm in}=5$ and $r_{\rm out}=80$ (in units of the mean atomic diameter) are fixed, so that no check for finite-size effects when $r_{\rm out}$ increases at a given $r_{\rm in}$ was provided in [\onlinecite{hentschel_eshelby}].

To reproduce this setup with a simple elasto-plastic model we consider a square lattice of linear size $L$ (using the lattice constant as the length unit) with periodic boundary conditions. We place at the center of the simulation box, taken as the origin, a large circular ``plastic defect'' $\mathcal D$ of diameter $\ell$ that mimics the Eshelby inclusion: see Fig.~\ref{fig_sketch-Eshelby}(b). The central defect contains blocks (or sites) that are on the verge of yielding and whose stress propagation axes are all with the same orientation (leading to a $\cos(4\theta)$ dependence with $\theta$ the angle with respect to the horizontal $x$-axis). At the beginning of the process, all central blocks yield with the same stress drop $\Delta\sigma_0$ and one follows the response of the medium by applying the kinetic rules of the elasto-plastic model.

The scalar nature of the variables in our elasto-plastic model is an (over)simplification of the problem which, however, makes sense if one considers, as already stated, that these variables are the XY component of the corresponding tensors and that, in addition, all the Eshelby stress propagators have the same orientation as the one associated with the large triggering defect. It is indeed likely that only such properly oriented Eshelby-like quadrupoles will first relax under the influence of the perturbation and will further trigger activity and generate quadrupoles with the same orientation. (Note that a similar assumption is made when describing a system under a uniform simple shear by a scalar elasto-plastic model; the relevant emergent Eshelby quadrupoles are not isotropically distributed but are rather oriented along the direction of the shear.\cite{EPM_review,picard_scalar,jagla_scalar,rossi_EPM}) 

More details on the elasto-plastic model, the setup, and the initial configurations are given in Appendix~\ref{app_EPM}. We consider several different preparations for the initial condition by varying the degree of disorder $R$ characterizing $P_0(\{\sigma_i\})$ or choosing marginal configurations generated by very slow steady-state uniform shear. In all cases, we obtain qualitatively similar results, within the resolution and scope of the present study.

\begin{figure}
\includegraphics[width=0.32\linewidth]{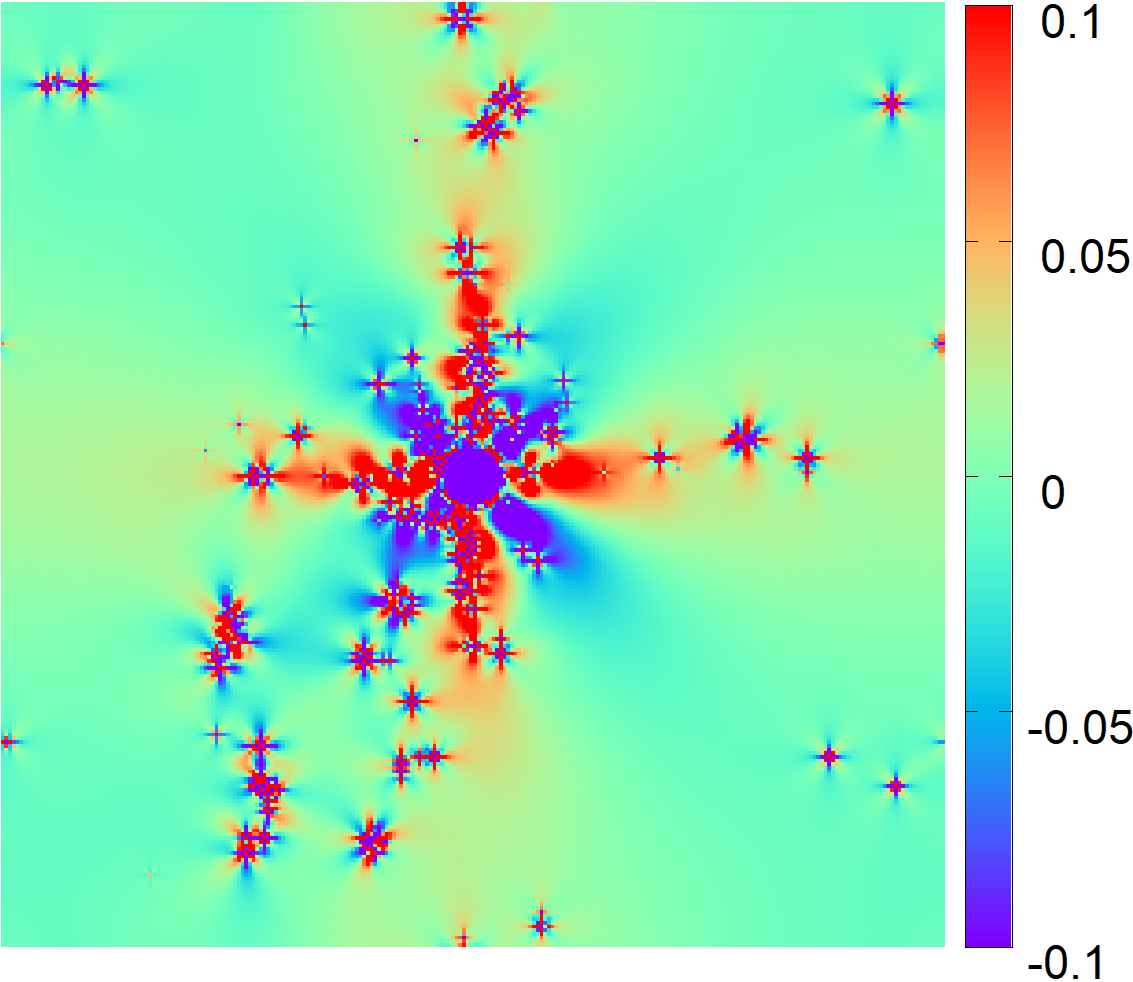}
\includegraphics[width=0.32\linewidth]{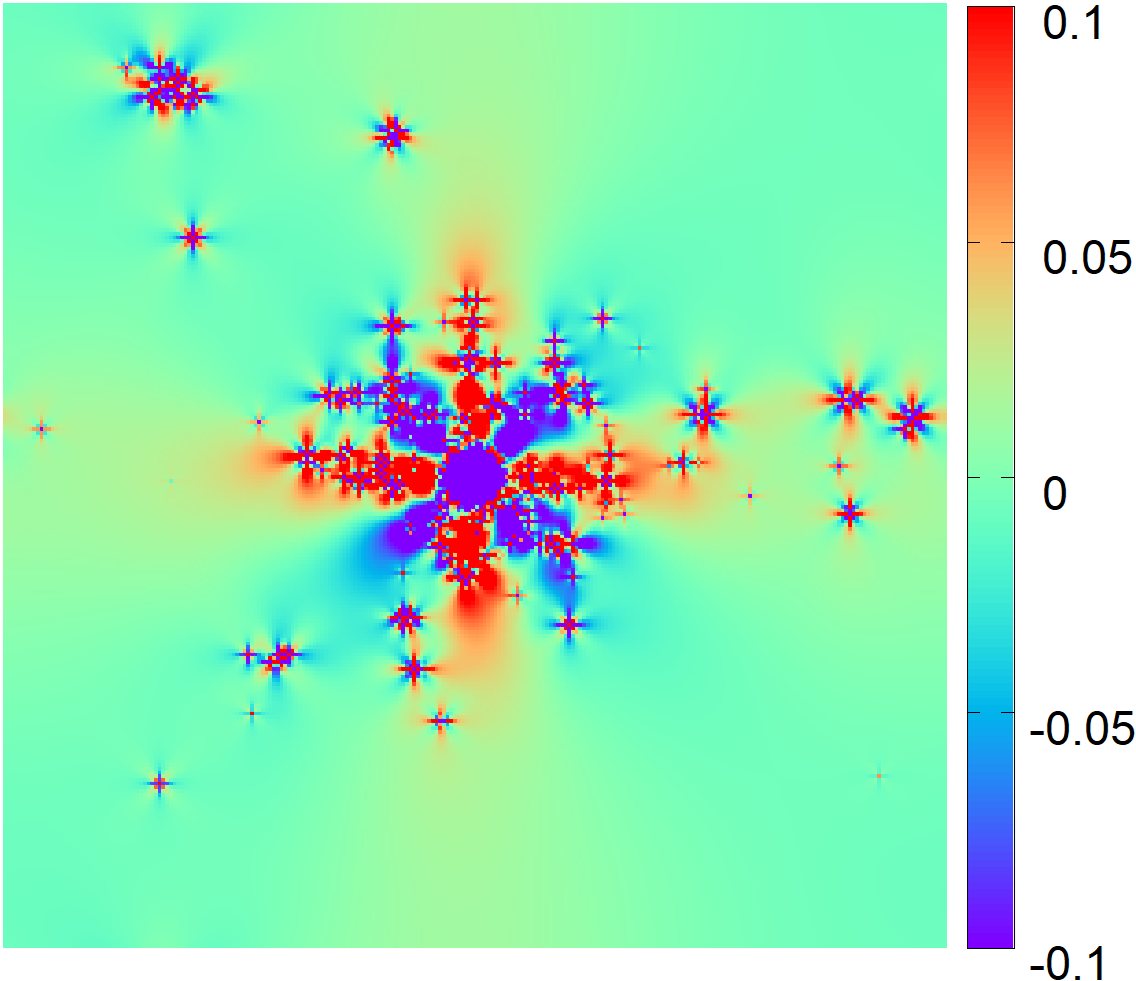}
\includegraphics[width=0.32\linewidth]{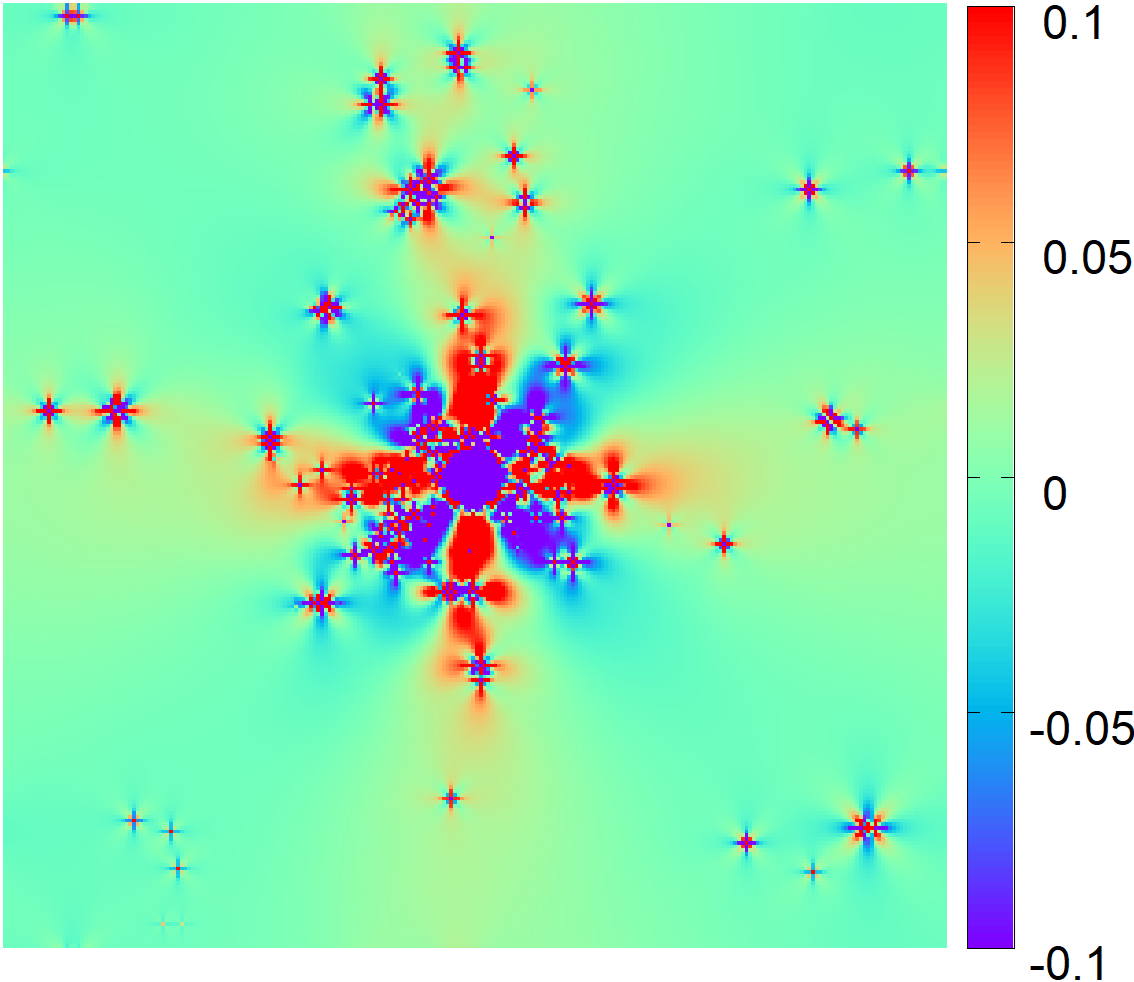}
\includegraphics[width=0.32\linewidth]{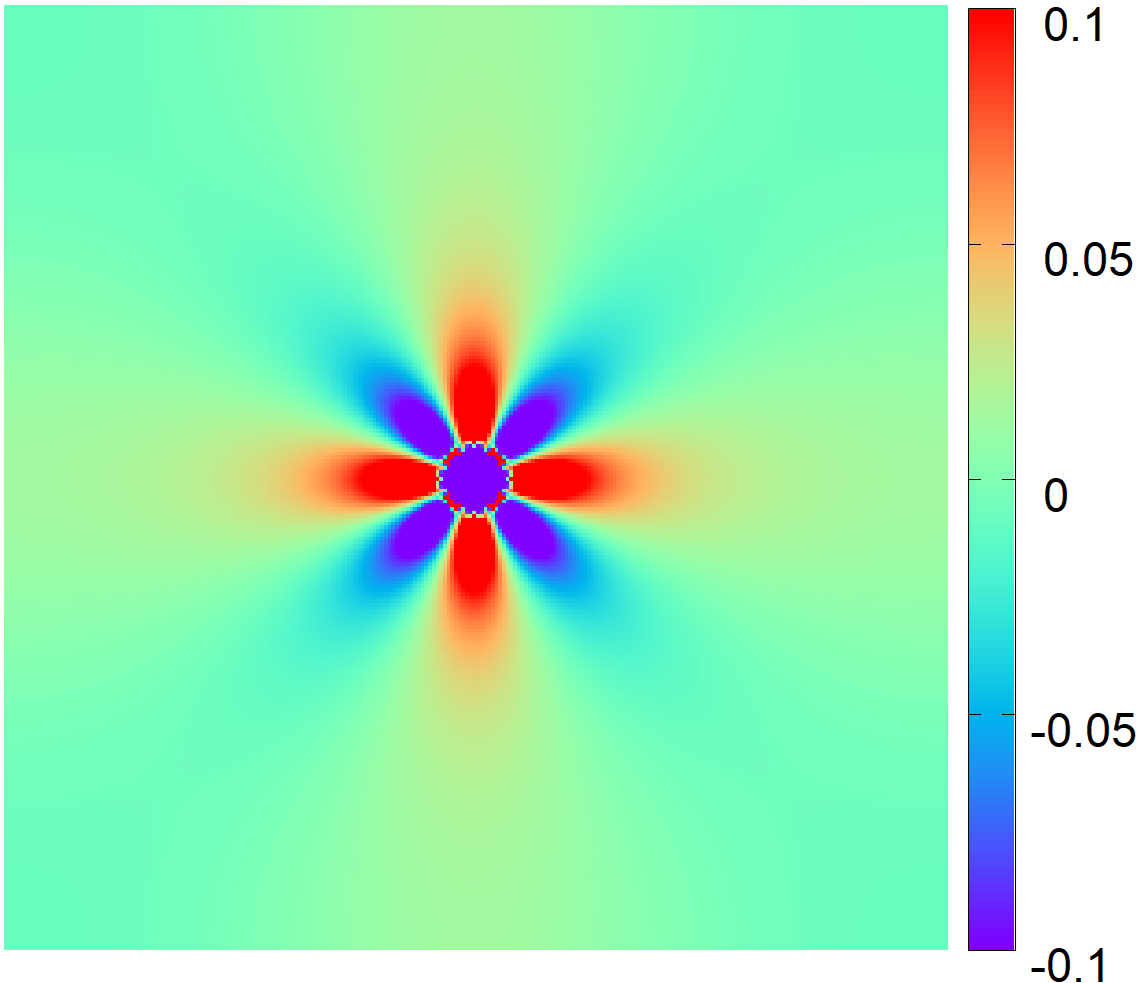}
\includegraphics[width=0.32\linewidth]{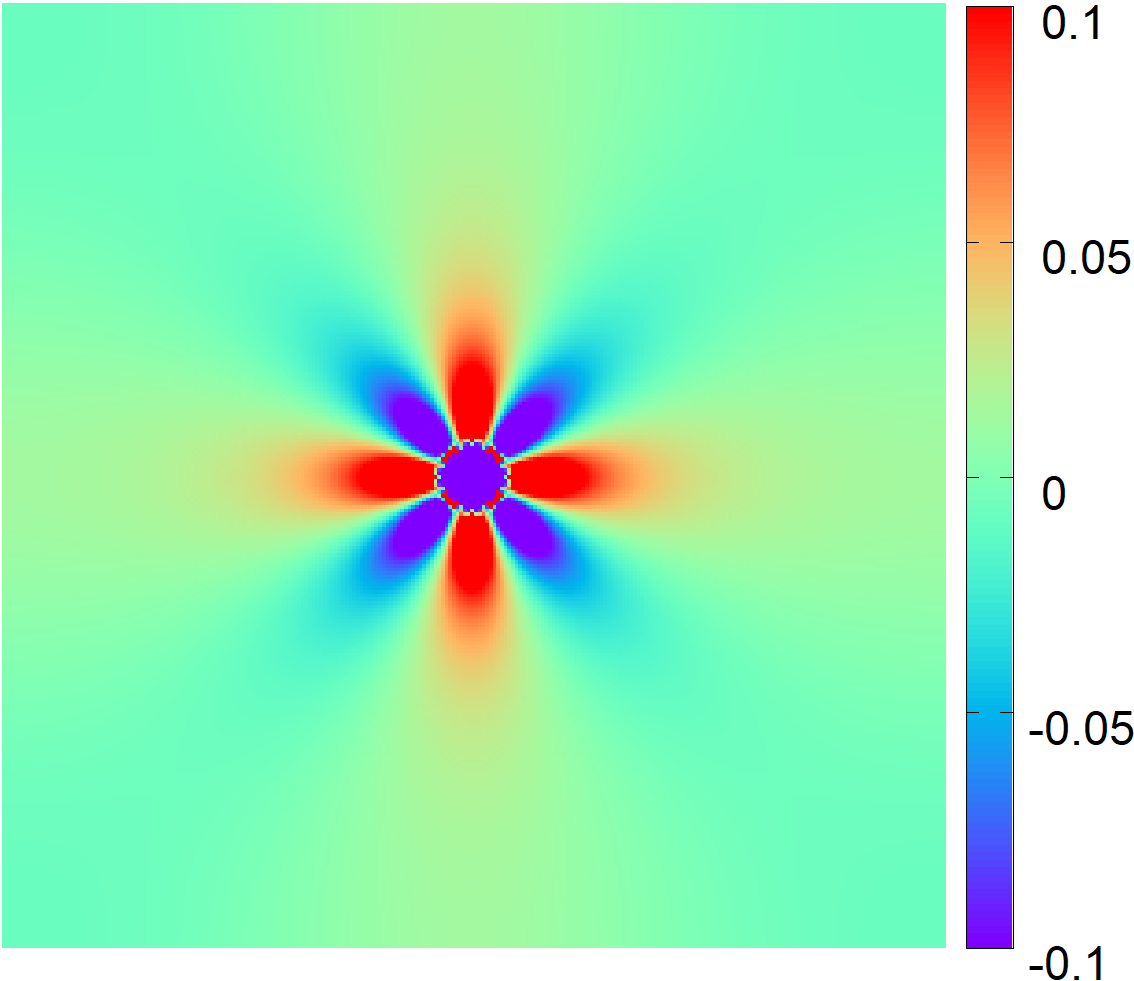}
\includegraphics[width=0.32\linewidth]{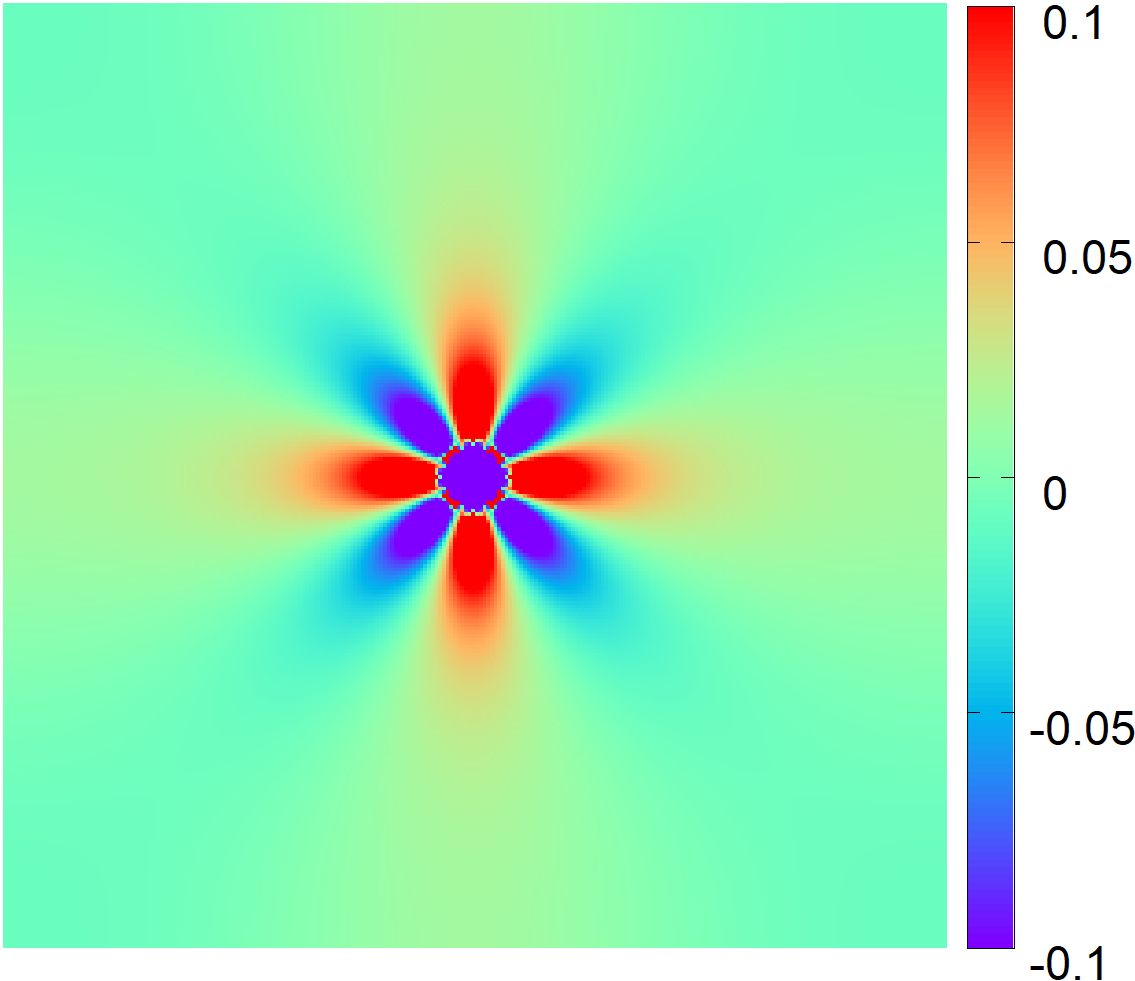}
\includegraphics[width=0.32\linewidth]{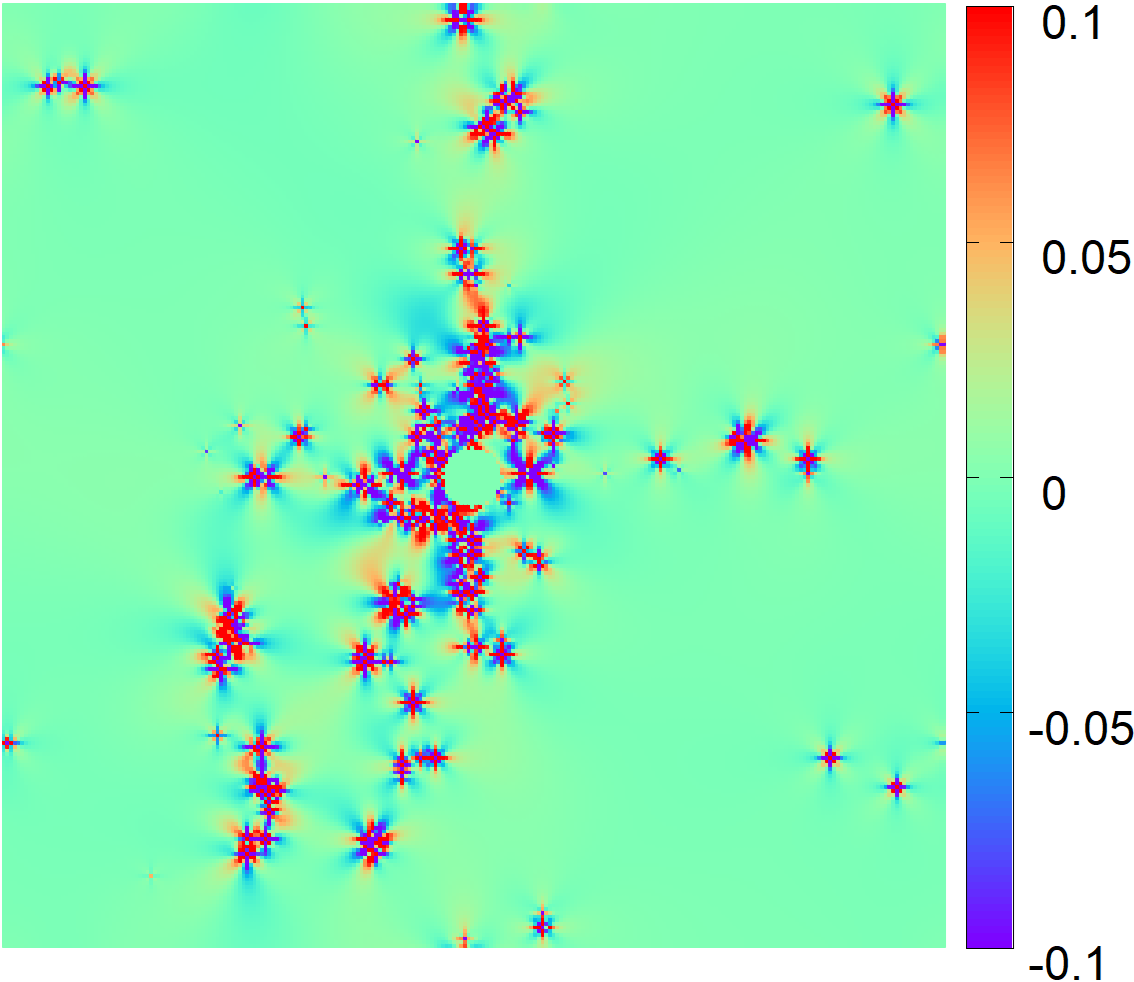}
\includegraphics[width=0.32\linewidth]{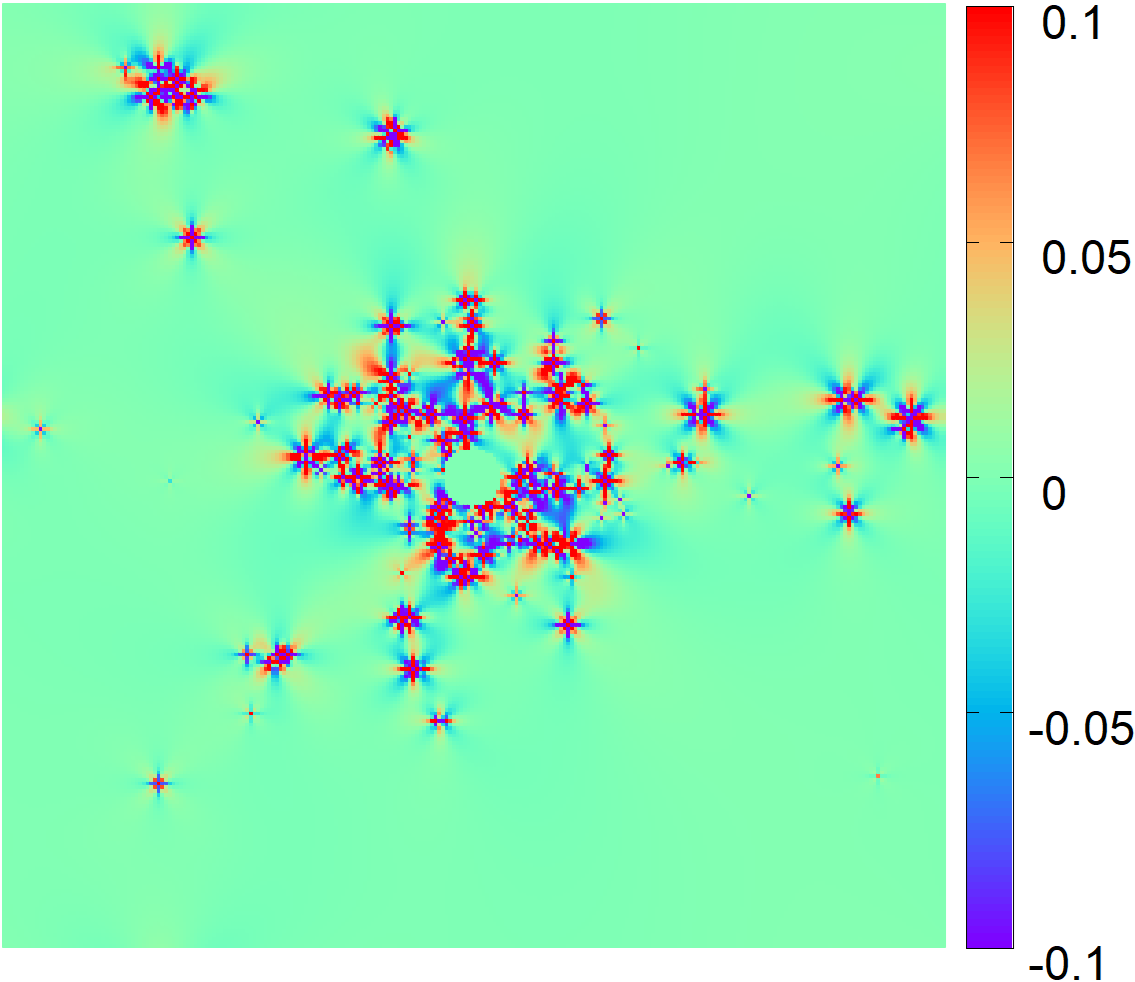}
\includegraphics[width=0.32\linewidth]{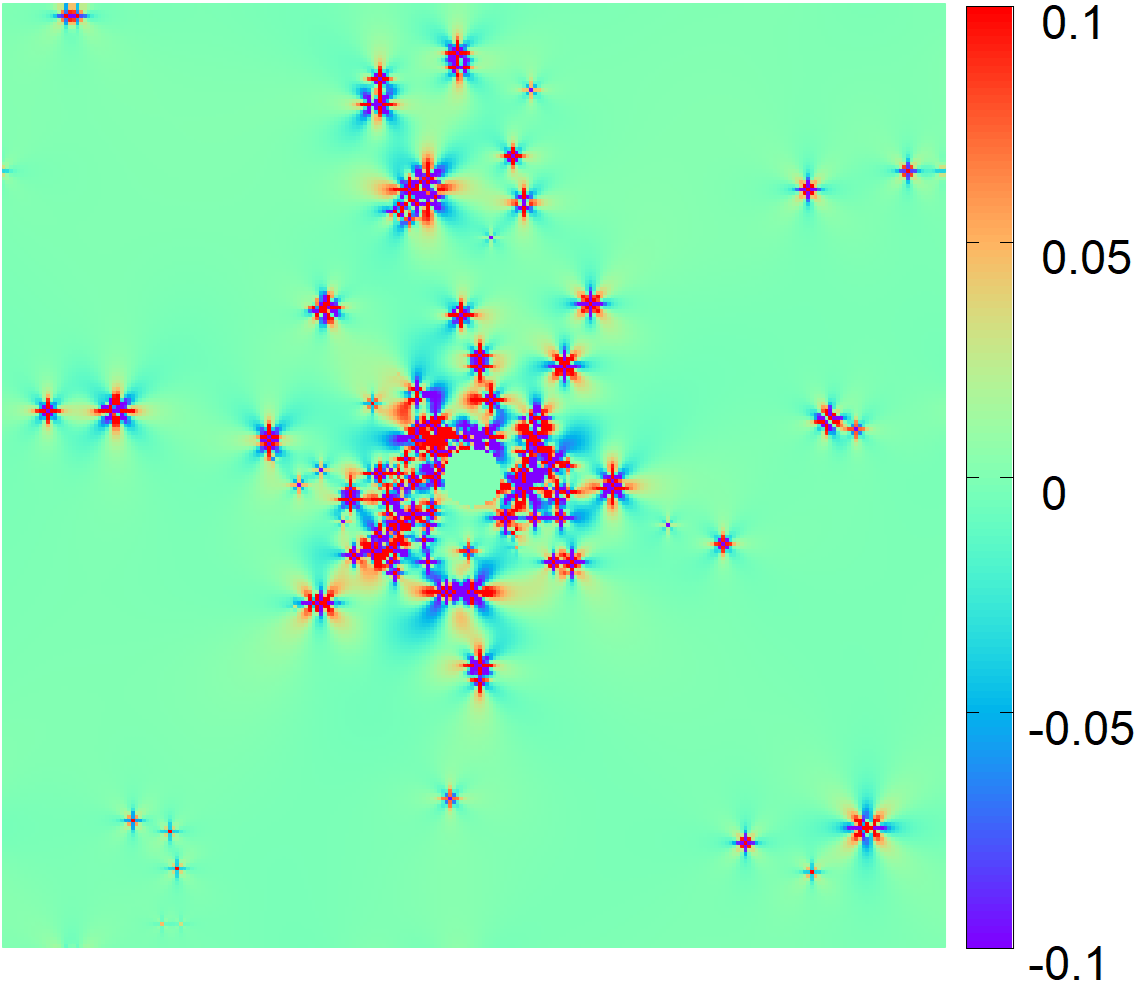}
\caption{
Elasto-plastic model: Stress-field difference induced by the initial Eshelby perturbation with a central defect of diameter $\ell=16$ and a system size $L=256$. Illustration for three different samples (left, middle, and right columns). The stress drop in the blocks forming the initial plastic defect is $\Delta\sigma_0=1$ and the shear modulus is taken as $\mu\equiv 1$. Top: full stress; middle: elastic component of the stress; bottom: plastic component of the stress. The initial plastic events associated with the central defect (shown as a blue core because the stress difference is $-\Delta \sigma_0=-1$) trigger additional plastic events (with aligned Eshelby-like quadrupoles) in the system. The color code is shown on the right of each panel. 
}
\label{fig_Eshelby_samples_large_defect}
\end{figure}

\begin{figure}
\includegraphics[width=0.32\linewidth]{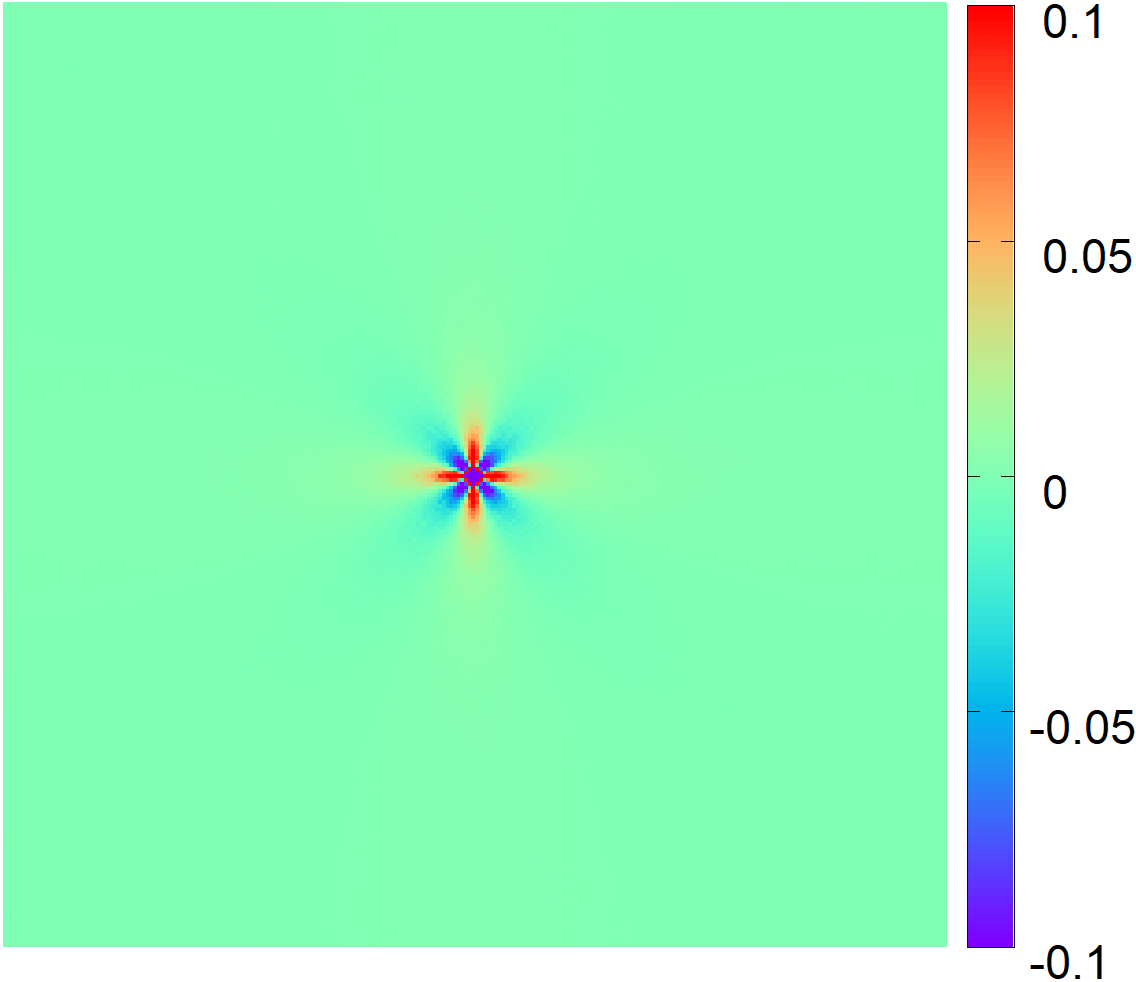}
\includegraphics[width=0.32\linewidth]{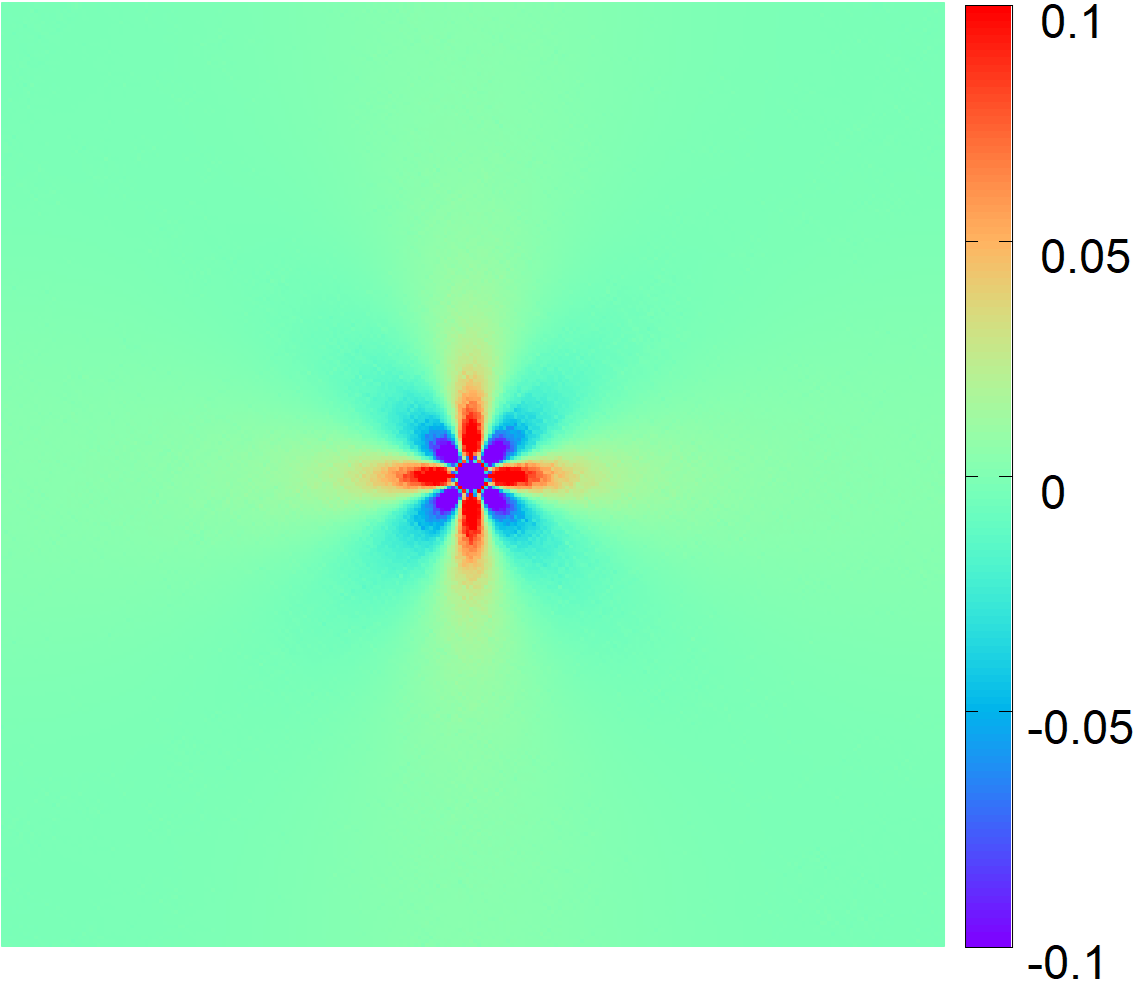}
\includegraphics[width=0.32\linewidth]{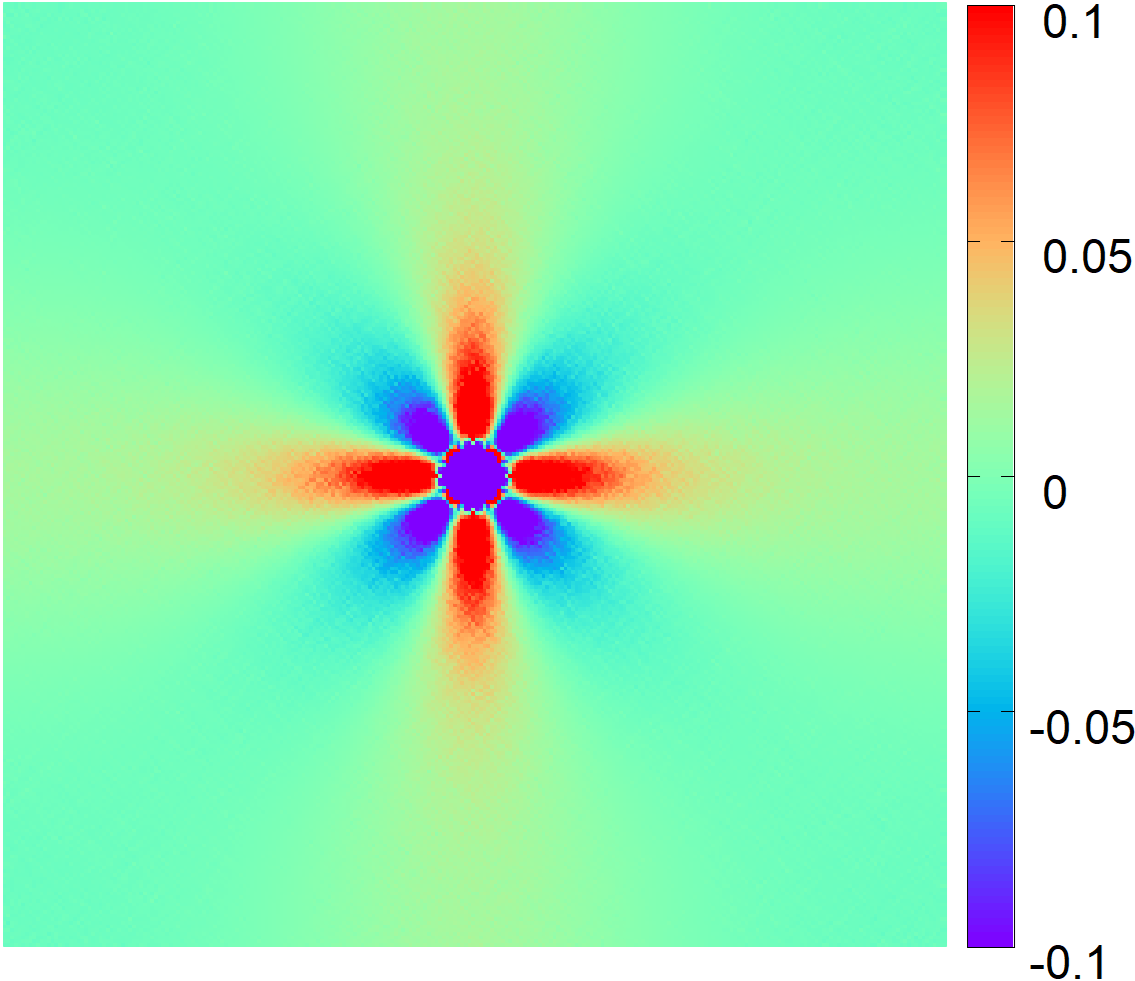}
\caption{
Elasto-plastic model: Stress-field difference induced by the initial Eshelby perturbation after averaging over 1000 independent samples for $\ell=4$ (left), $\ell= 8$ (middle), and $\ell = 16$ (right). The stress drop in the blocks forming the initial plastic defect (shown as a blue central core) is $\Delta\sigma_0=1$, $\mu=1$, and the system size is $L=256$. The color code is shown on the right of each panel. 
}
\label{fig_stress_field_average}
\end{figure}

\begin{figure}
\includegraphics[width=.8\linewidth]{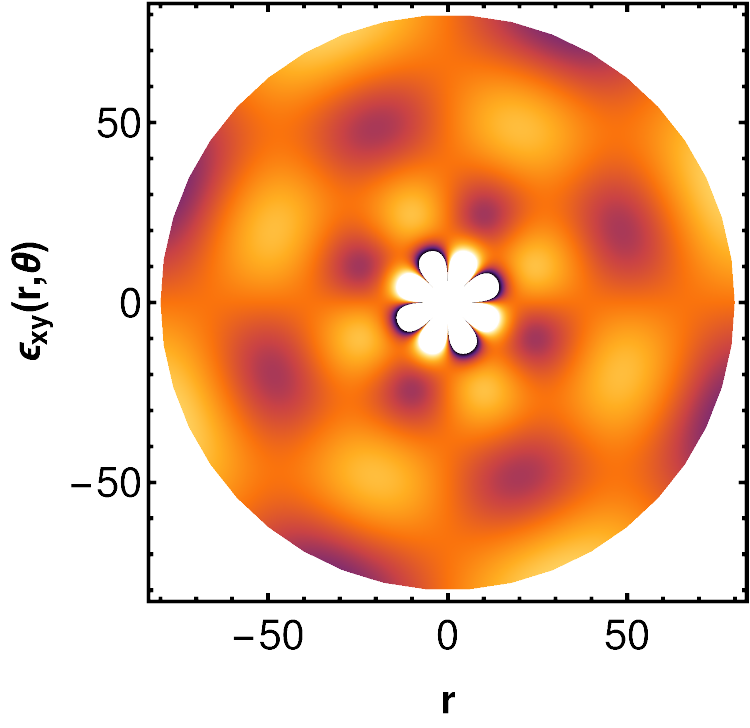}
\caption{Atomic simulation: Sample-averaged XY component of the strain field for the Eshelby problem with $r_{\rm in}=5$, $r_{\rm out}=80$, and $\delta=1.05$ (see Fig.~\ref{fig_sketch-Eshelby}(a)). Notice the overall $\cos(4\theta)$ symmetry (here, up to a phase). The XY strain field is obtained by combining simulation results for the displacement field  in a poorly annealed glass and an analytical description of anomalous elasticity.\cite{hentschel_eshelby,avanish-eshelby} (courtesy of  Avanish Kumar and Itamar Procaccia).
}
\label{fig_avanish_strain-map}
\end{figure}

\subsection{Quadrupolar Eshelby-like singularities and their density}
\label{sub_quadrupoles}

We first show some results for the (XY) stress field induced by the central Eshelby perturbation. It is obtained as the difference between the final stress configuration after the system has relaxed and the initial configuration prior to the perturbation in the region outside the imposed central defect:
\begin{equation}
\label{eq_local-deltasigma_difference}
\Delta\sigma_i^{\rm f/i}=\sigma_i^{\rm final}-\sigma_i^{\rm initial}, \; \forall i \notin \mathcal D,
\end{equation}
where it is recalled that $\mathcal D$ is the central defect. Plastic events in the form of quadrupolar Eshelby-like singularities are clearly visible in the samples displayed in Fig.~\ref{fig_Eshelby_samples_large_defect} and they appear nonuniformly distributed in each sample. 
When averaged over a large number of samples ($1000$ in the present case with $\ell=4$, $8$, $16$ and $L=256$), as shown in Fig.~\ref{fig_stress_field_average}, one recovers the overall $\cos(4\theta)$ symmetry of the stress-field change generated by an Eshelby inclusion, just like the XY strain field obtained in the atomic simulation of the Eshelby problem:\cite{hentschel_eshelby,avanish-eshelby} see Fig.~\ref{fig_avanish_strain-map} for illustration.

We now consider  the density of plastic events and the finite-size effects at fixed defect size $\ell$ discussed in Sec.~\ref{sec_finite-size}. To do so, we monitor the plastic activity at each site $i \notin \mathcal D$ after the initial Eshelby-like perturbation. We define an associated field as
\begin{equation}
n({\bf r})=\sum_{i\notin \mathcal D} n_i \delta^{(d)}({\bf r}-{\bf r}_i),
\label{eq:n_r}
\end{equation}
where $n_i=1$ or $0$ depending on whether site $i$ has undergone at least one plastic event or not. Up to a stress drop $\Delta\sigma_i$ which fluctuates around a nonzero mean, $n({\bf r})$ is essentially the quadrupole field in a given sample: see Eq.~(\ref{eq_EPM_quadrupole}). (Note that sites can in principle undergo several plastic events during the process but we have found that this rather rarely occurs and does not change the conclusions: see Appendix~\ref{sec:multiple_yielding}.)

We then compute the sample-averaged number of plastically active sites in a circular domain of radius $r$ centered on the origin (but excluding the initial Eshelby defect $\mathcal D$),
\begin{equation}
\overline N(r)=\int_{\ell/2}^r dr' r' \int_0^{2\pi} d\theta' \overline n(\bf r'),
\label{eq:bar_N}
\end{equation}
where ${\bf r}'=(r',\theta')$ in polar coordinates and $\overline{(\cdots)}$ is the average over samples. The numerical computation of Eq.~(\ref{eq:bar_N}) is explained in Appendix~\ref{sec:numerical_computation}.
 
\begin{figure}
\includegraphics[width=.98\linewidth]{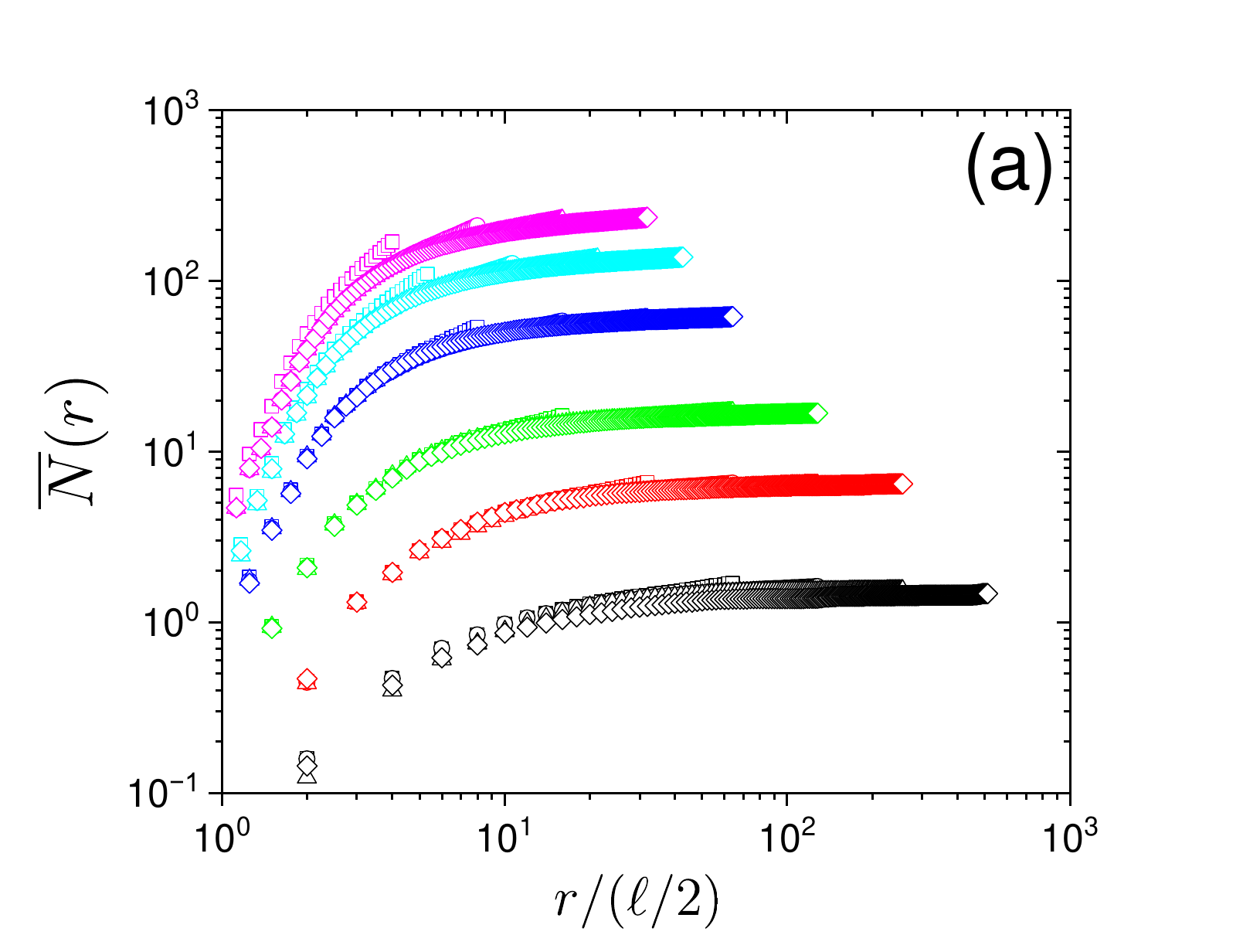}
\includegraphics[width=.98\linewidth]{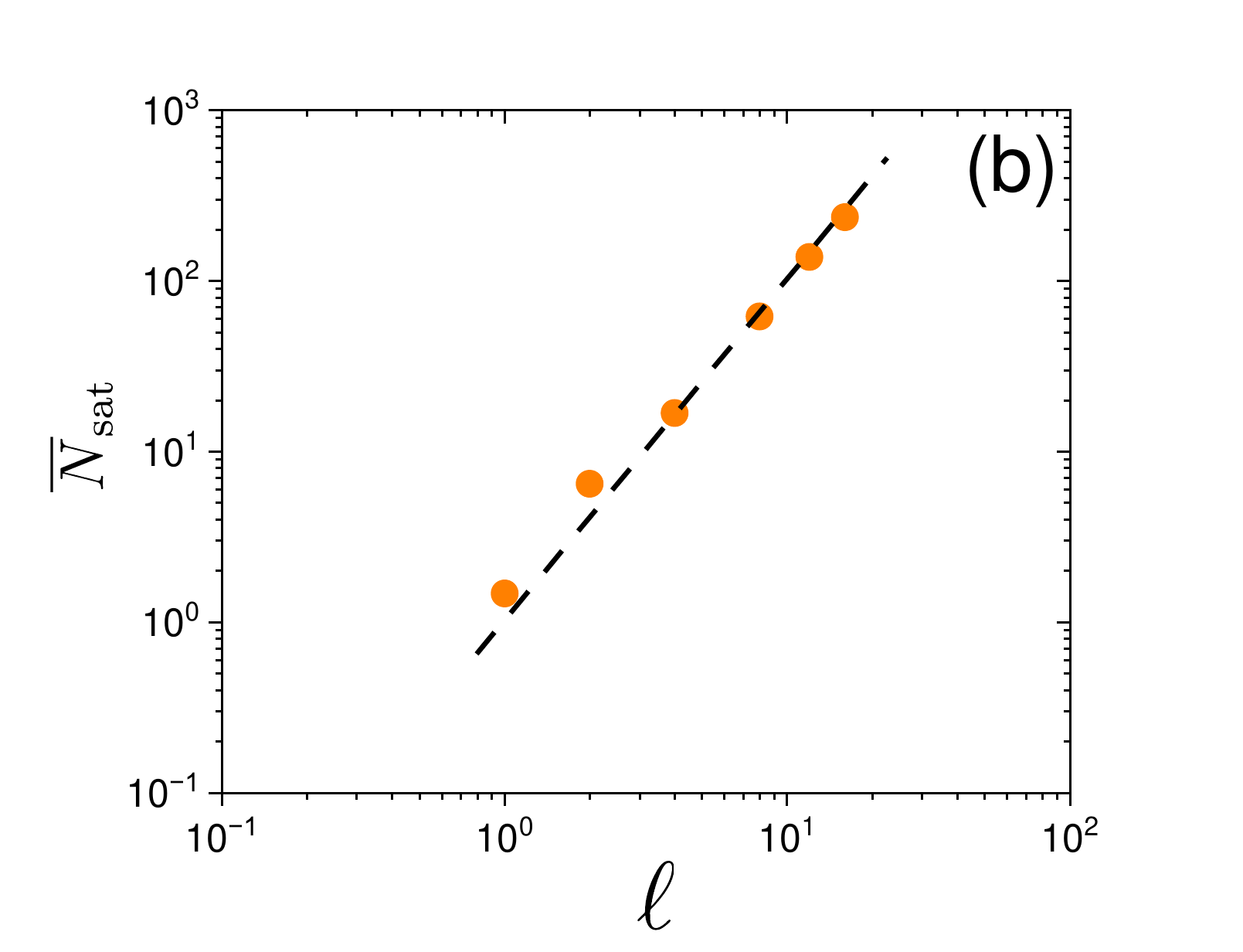}
\includegraphics[width=.98\linewidth]{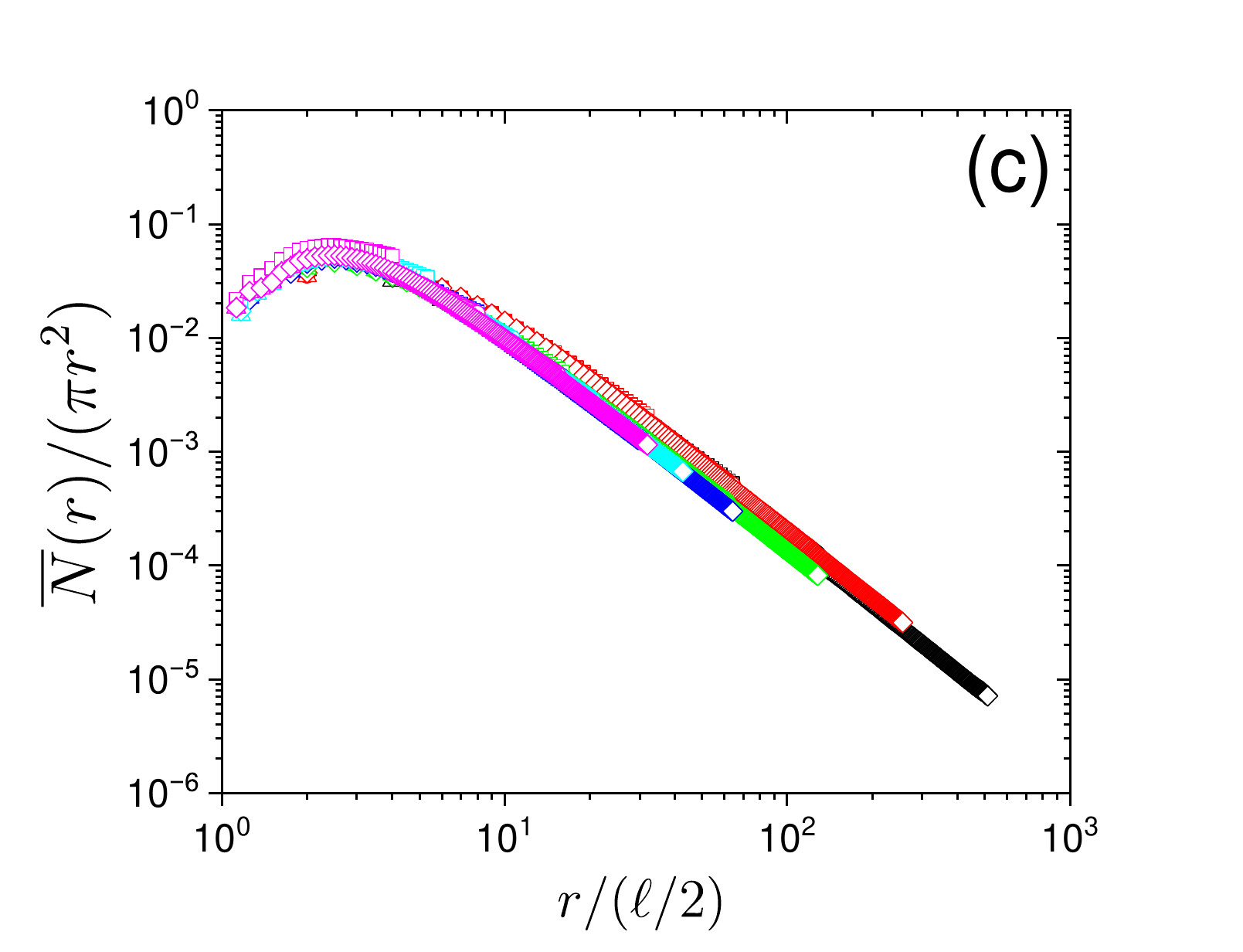}
\caption{(a) Averaged number of plastically active sites $\overline N(r)$ in an annulus (centered on the origin) between the boundary of the central defect at $\ell/2$ and $r$ for several values of the initial defect size, $\ell=1$ (black), 2 (red), 4 (green), 8 (blue), 12 (light blue), and 16 (pink) (from bottom to top), and four different system sizes, $L=64$ (squares), $L=128$ (circles), $L=256$ (triangles), and $L=512$ (diamonds). The distance $r$ is normalized by the defect radius $\ell/2$. 
Notice the saturation toward a value $\overline{N}_{\rm sat}$ that is rather independent of the system size but grows with $\ell$. (b) The saturation value $\overline{N}_{\rm sat}$ grows as $\overline{N}_{\rm sat} \sim \ell^2$ (dashed line).
(c) Averaged density of plastically active sites $\overline{N}(r)/(4\pi r^2)$ versus $r/(\ell/2)$.
}
\label{fig_number_plastic-events_Eshelby}
\end{figure}

We illustrate the outcome for several values of the defect size $\ell$ and of the system size $L$ and for a perturbation amplitude $\mu\Delta\sigma_0=1$. This is plotted in Fig.~\ref{fig_number_plastic-events_Eshelby}(a) where the distance $r$ is normalized by the radius of 
the central defect, $\ell/2$. One can see that at a distance that roughly scales with $\ell$ the curves all saturate to a value $\overline{N}_{\rm sat}(\ell,L)$ that essentially does not depend on the system size $L$ (for $L/\ell$ large enough). However, the saturation value grows with $\ell$. More precisely, by displaying $\overline{N}_{\rm sat}$ versus $\ell$ on a log-log plot we find that it grows as $\ell^2$: see Fig.~\ref{fig_number_plastic-events_Eshelby}(b).
This shows that a nonzero density of plastically active sites, or equivalently of quadrupoles, can exist over a distance that scales with the size of the applied perturbation $\ell$ but that the density vanishes in the bulk of the system in the thermodynamic limit: This density indeed goes to zero when the system size $L$ goes to infinity as $\overline N_{\rm sat}(\ell)/L^2\sim (\ell/L)^2$. As argued in Sec.~\ref{sec_finite-size}, this result is likely to be general and only depends on gross features of the problem that are captured by the elasto-plastic model in spite of its limitations. 
 
Finally, note that even in the region where it is nonzero the density of plastic events remains small. This is illustrated in Fig.~\ref{fig_number_plastic-events_Eshelby}(c) where we display $\overline{N}(r)/(4\pi r^2)$ versus $r/(\ell/2)$. For instance, for $r$ between 3 and 10 times $\ell/2$, the density decreases from  $5\%$ and $1\%$.

\subsection{Renormalization of elasticity versus dipole screening}
\label{sub_results}

As expected, quadrupoles are generated in the Eshelby problem described by an elasto-plastic model, at least in a region whose size is set by that of the initial perturbation $\ell$, and they arrange in a nonuniform way in any given sample. But does this actually lead to anomalous elasticity in the region where a nonzero density of quadrupoles is found?

To address this question, we introduce the sample-averaged stress field that is produced by the full (elastic and plastic) response of the medium to the initial Eshelby perturbation,
\begin{equation}
    \overline{\Delta\sigma}({\bf r})=\sum_{i \notin \mathcal D} \overline{\Delta\sigma_i}^{\rm f/i}\delta^{(d)}({\bf r}-{\bf r}_i),
    \label{eq:Delta_sigma_def}
\end{equation} 
with $\Delta\sigma_i^{\rm f/i}$ defined in Eq.~(\ref{eq_local-deltasigma_difference}). We also consider its component due to the purely elastic response,
\begin{equation}
\Delta\sigma^{\rm elastic}({\bf r})=\sum_{i \notin \mathcal D} \Delta\sigma_i^{\rm f/i,elastic}\delta^{(d)}({\bf r}-{\bf r}_i) 
\end{equation}
with
\begin{equation}
\Delta\sigma_i^{\rm f/i,elastic}=\sum_{j\in\mathcal D}G_{ij}^{\rm Eshelby} \mu\Delta\sigma_0,
\end{equation}
which does not depend on the sample, and the complementary component due to the plastic response, 
\begin{equation}
\label{eq_plastic_stress-field}
\overline{\Delta\sigma}^{\rm plastic}({\bf r})=\overline{\Delta\sigma}({\bf r})-\Delta\sigma^{\rm elastic}({\bf r}).
\end{equation}
Note that the plastic response so defined corresponds to the difference between the final and initial states, as in the computer simulations and in the continuum theory.\cite{hentschel_eshelby} It now takes into account the possibility that a site has been plastically active several times during the process but, as already mentioned and discussed in Appendix~\ref{sec:multiple_yielding}, this effect is anyhow small.

As seen in Fig.~\ref{fig_stress_field_average}, the sample-averaged stress field has the expected symmetry of the Eshelby problem (and so does its elastic component). We then focus on the radial components that can be obtained by some averaging over the polar angle, {\it e.g.},
\begin{equation}
\overline{\Delta\sigma}(r) = \frac 1{\pi}\int_0^{2\pi} d\theta \cos(4\theta) \overline{\Delta\sigma}({\bf r})     
\label{eq:Delta_sigma_average}
\end{equation}
and similarly for the elastic and plastic components. (The numerical computation of $\overline{\Delta\sigma}(r)$ is explained in Appendix~\ref{sec:numerical_computation}.)

\begin{figure*}
\includegraphics[width=0.32\linewidth]{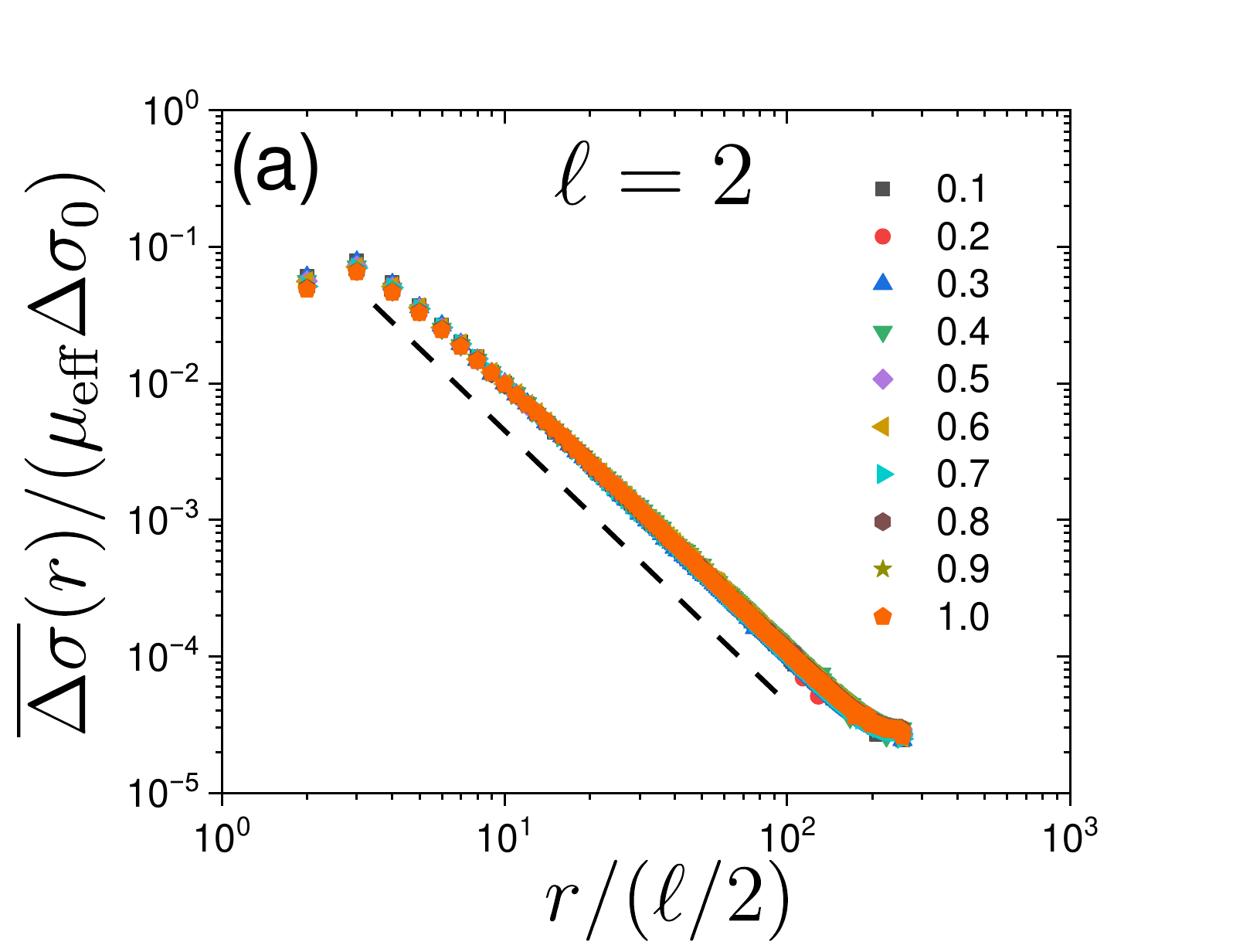}
\includegraphics[width=0.32\linewidth]{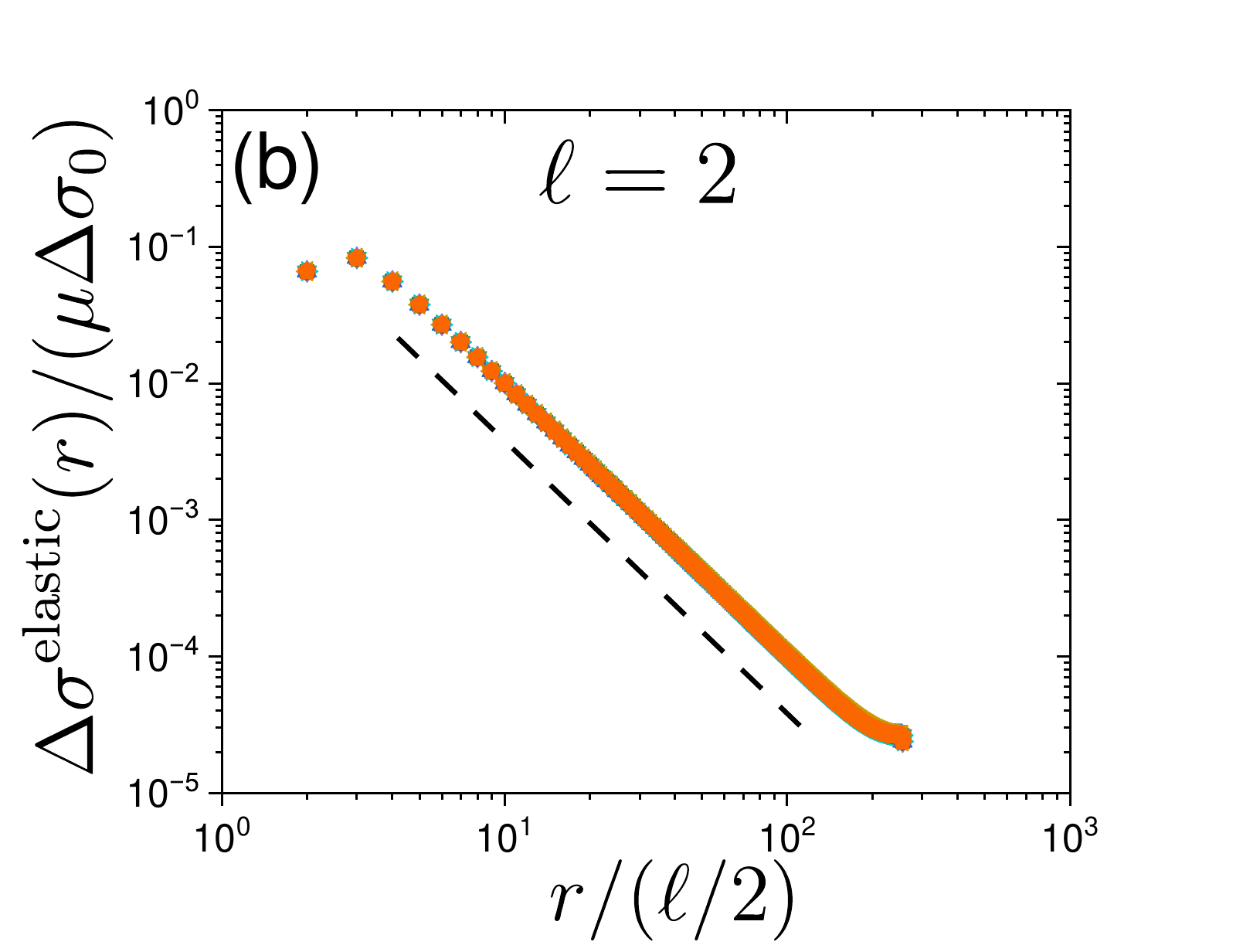}
\includegraphics[width=0.32\linewidth]{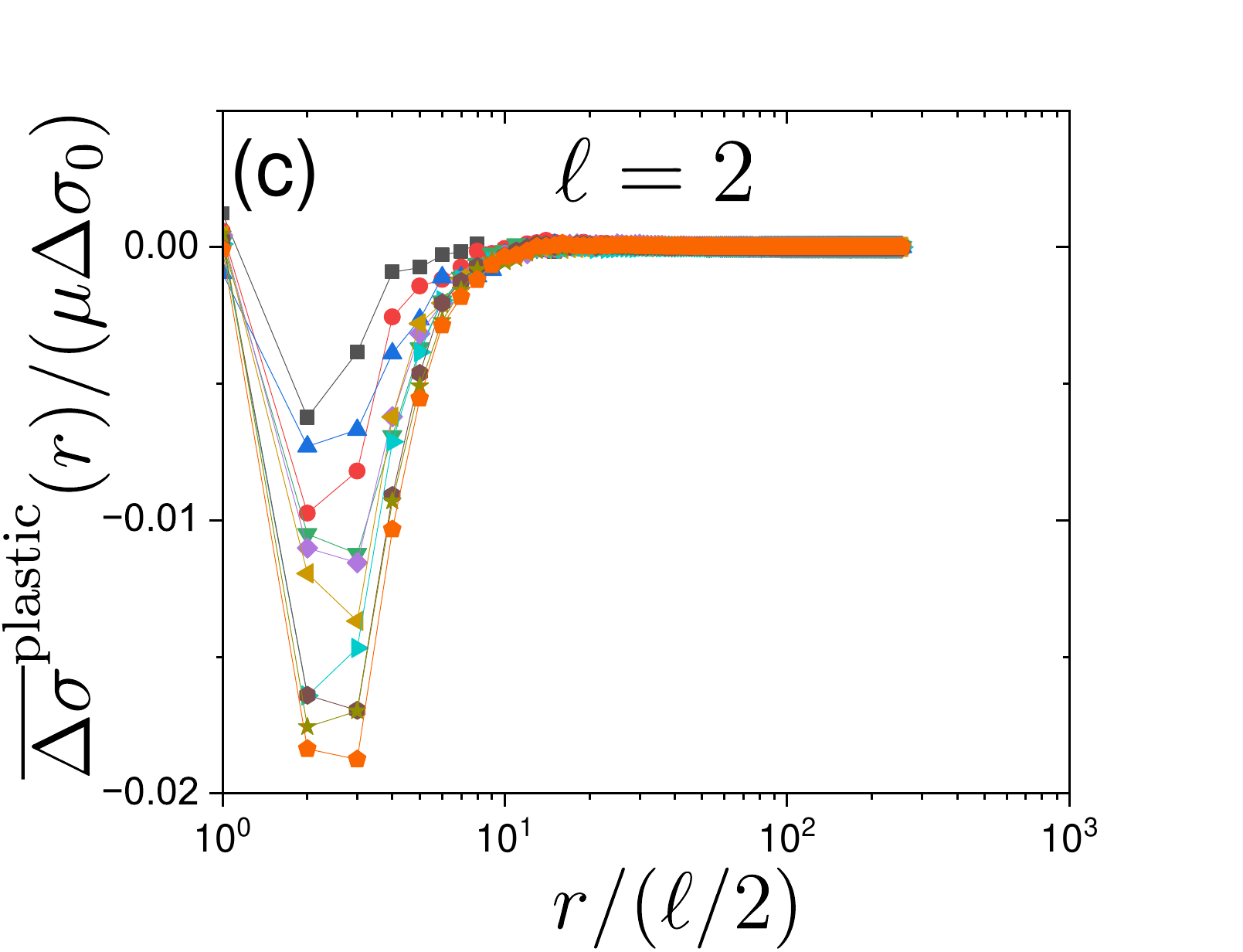}
\includegraphics[width=0.32\linewidth]{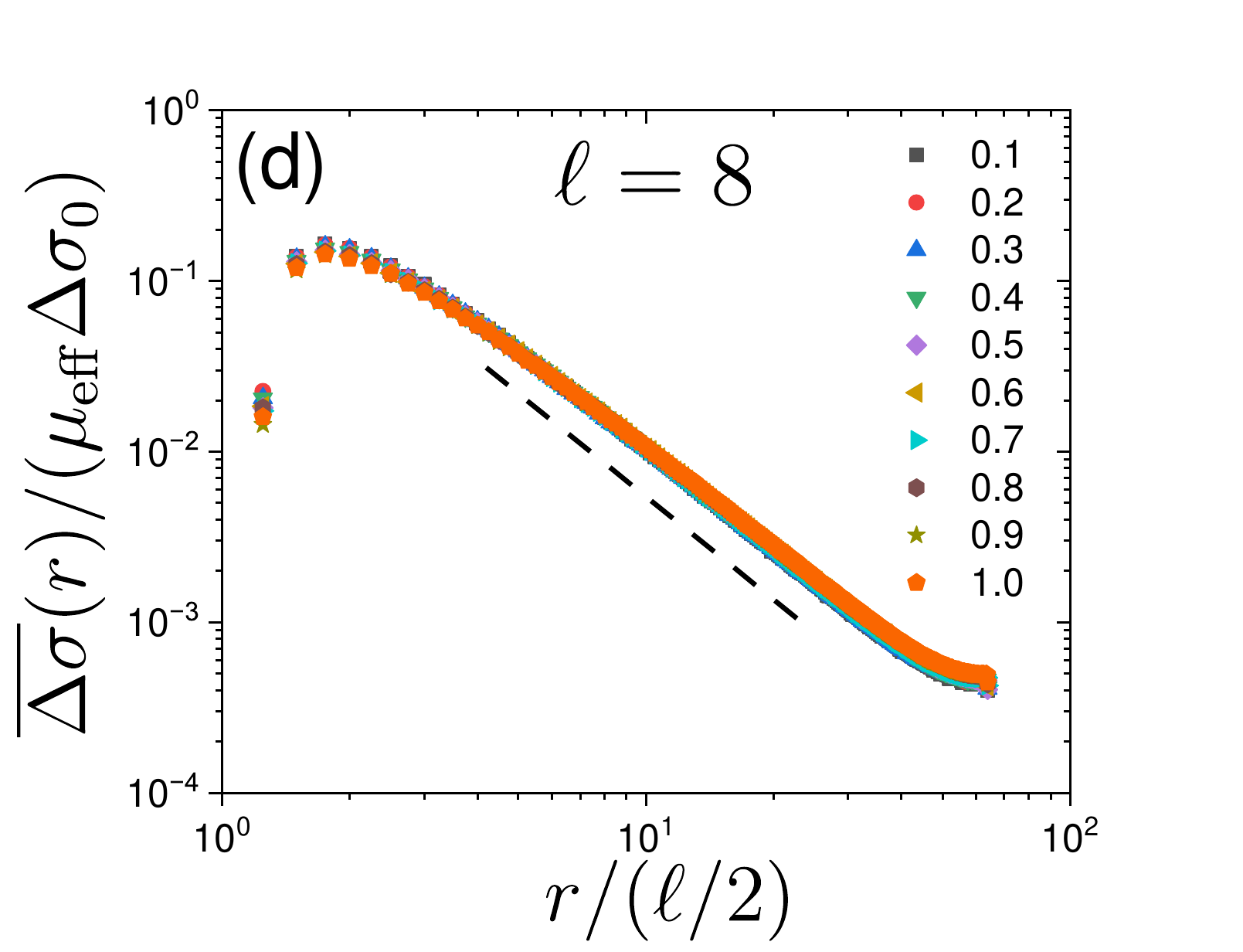}
\includegraphics[width=0.32\linewidth]{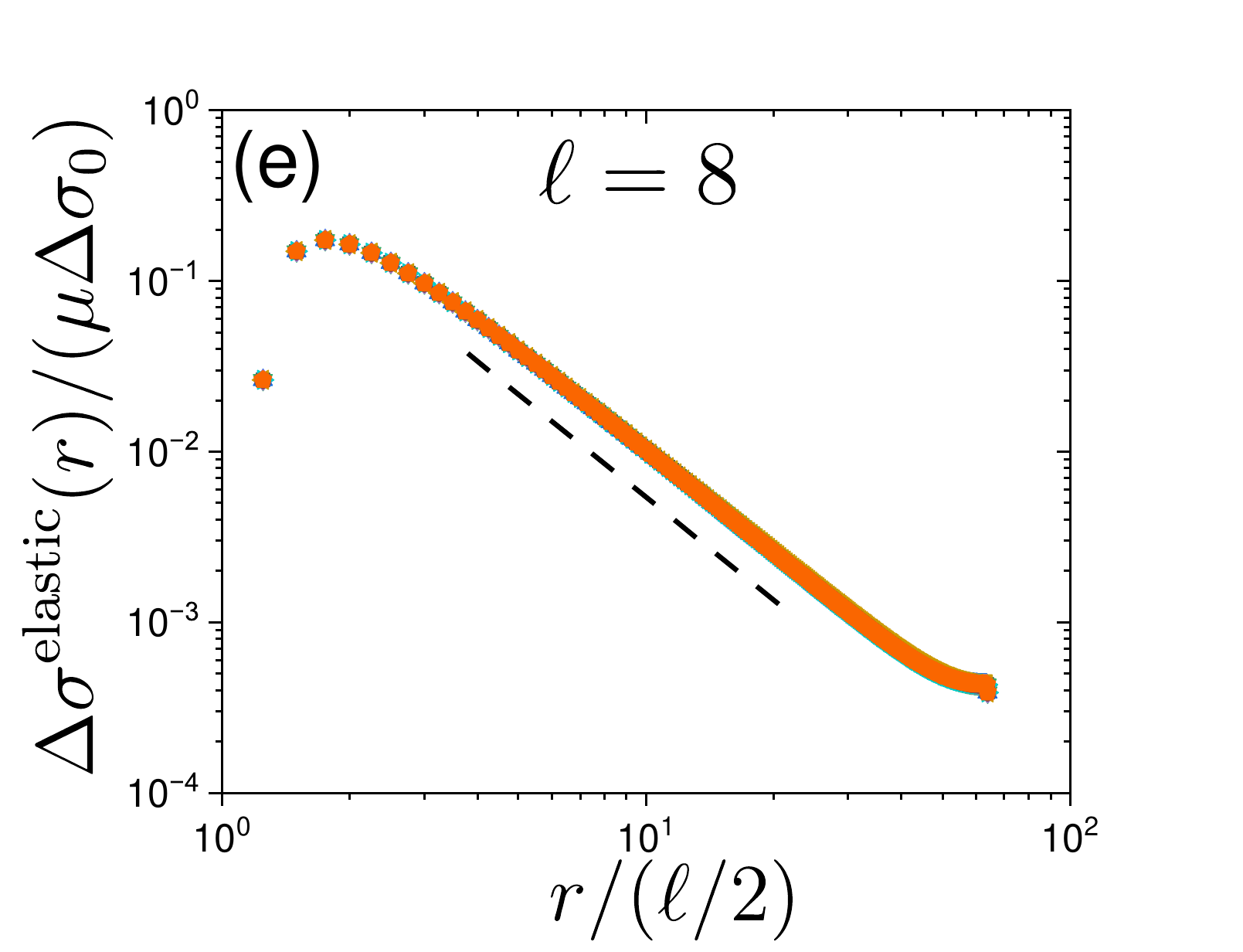}
\includegraphics[width=0.32\linewidth]{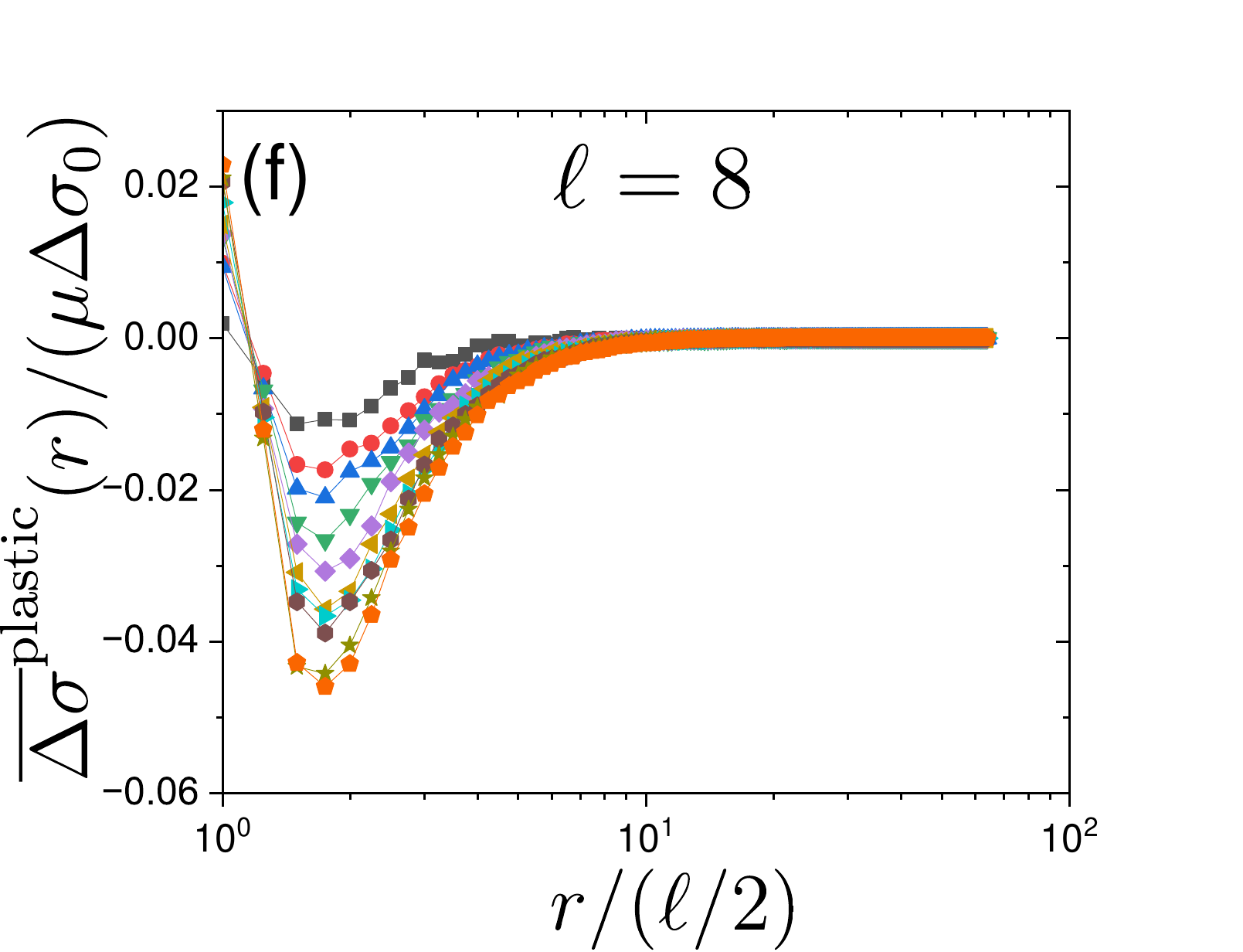}
\includegraphics[width=0.32\linewidth]{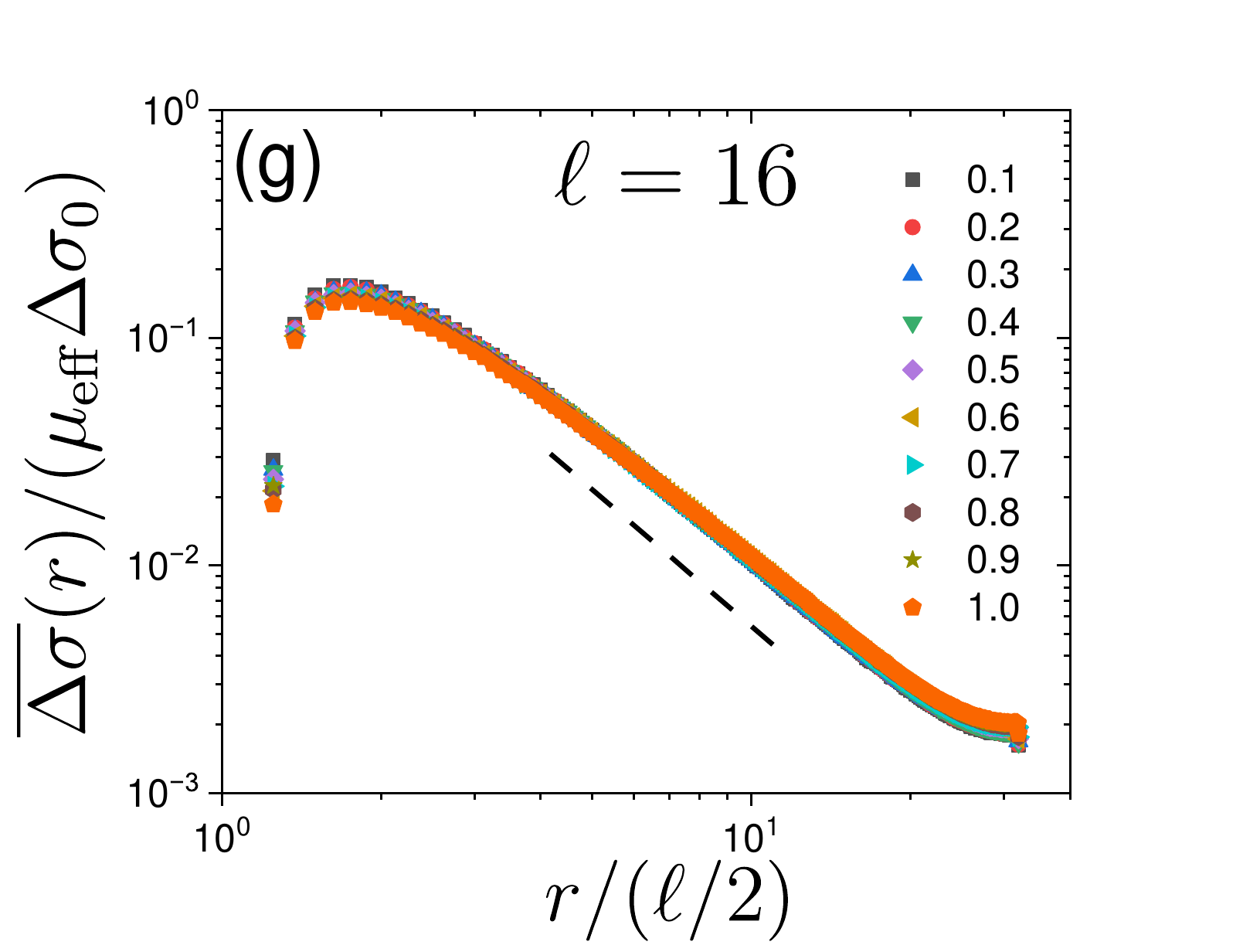}
\includegraphics[width=0.32\linewidth]{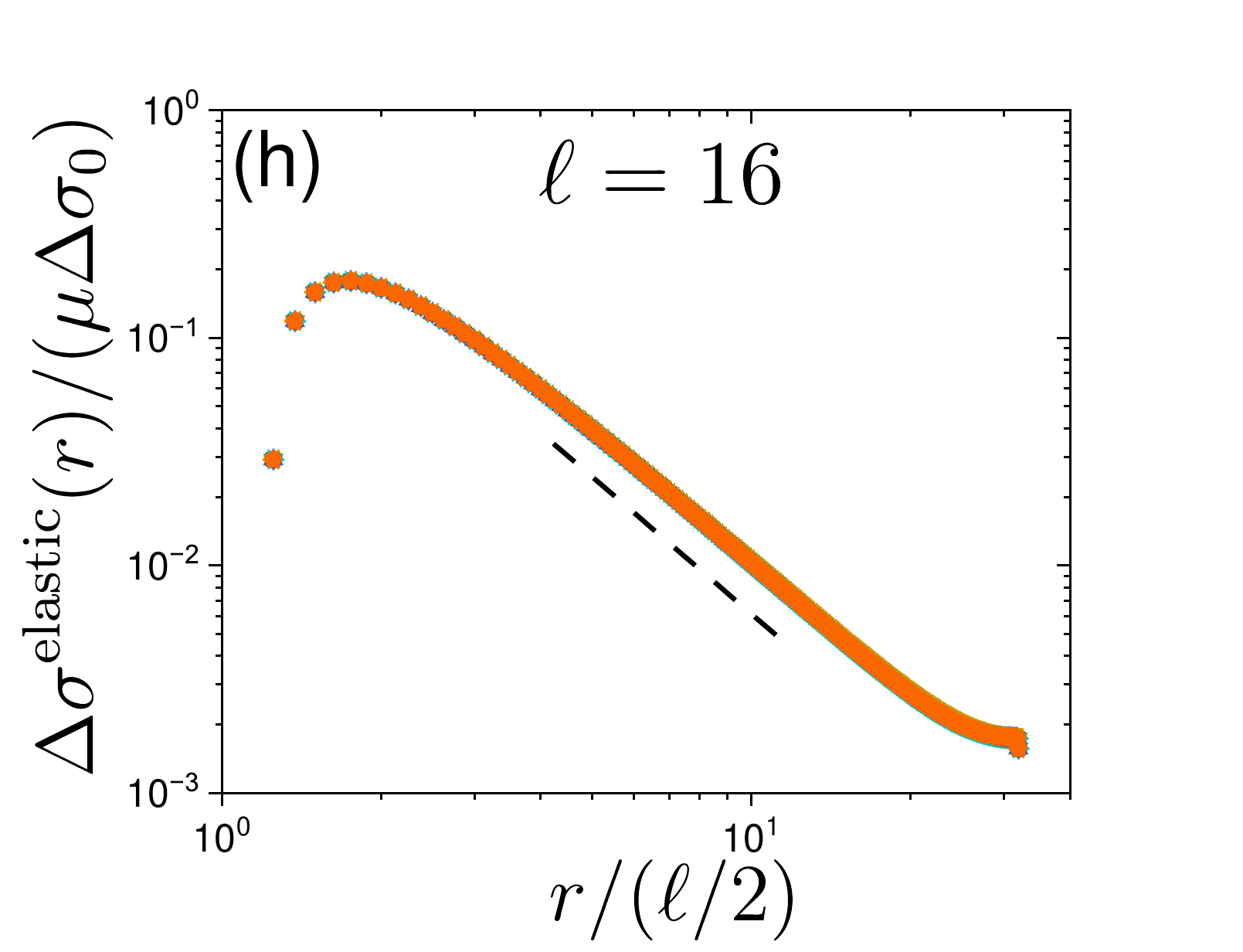}
\includegraphics[width=0.32\linewidth]{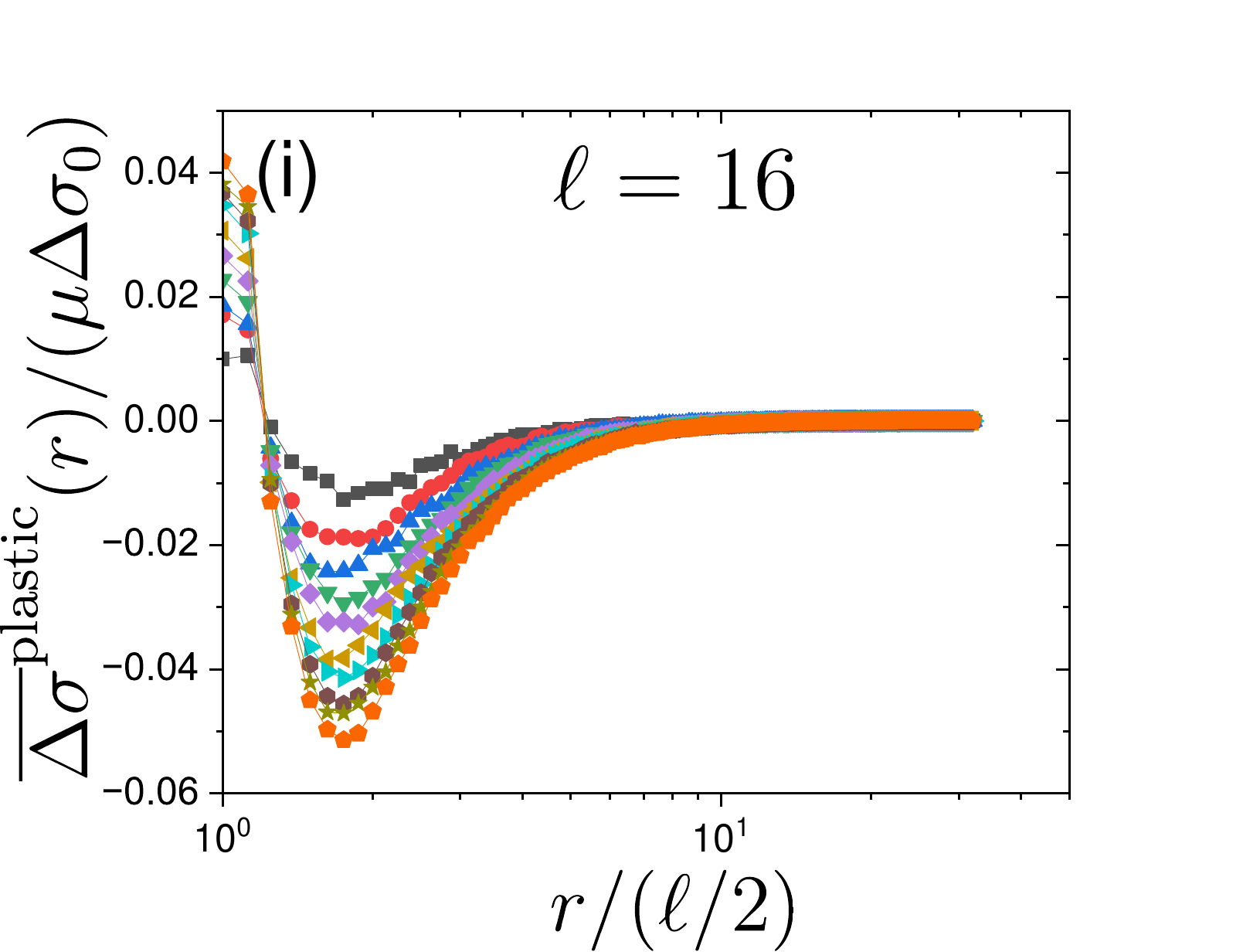}
\caption{Full angle-averaged induced stress $\overline{\Delta\sigma}(r)$ and its elastic and plastic components, normalized by the amplitude of the perturbation $\mu \Delta\sigma_0$, for several values of the defect size $\ell$ and of $\mu \Delta\sigma_0$. The system size is $L=512$. From top to bottom: $\ell=2$, 8, and 16. The dashed line indicate the expected slope of $-2$ corresponding to the leading term in Eqs.~(\ref{eq_Delta_sigma_elastic_continuum}) and (\ref{eq_Delta_sigma_full_continuum}). 
}
\label{fig_renormalization}
\end{figure*}

\begin{figure}
\includegraphics[width=0.98\linewidth]{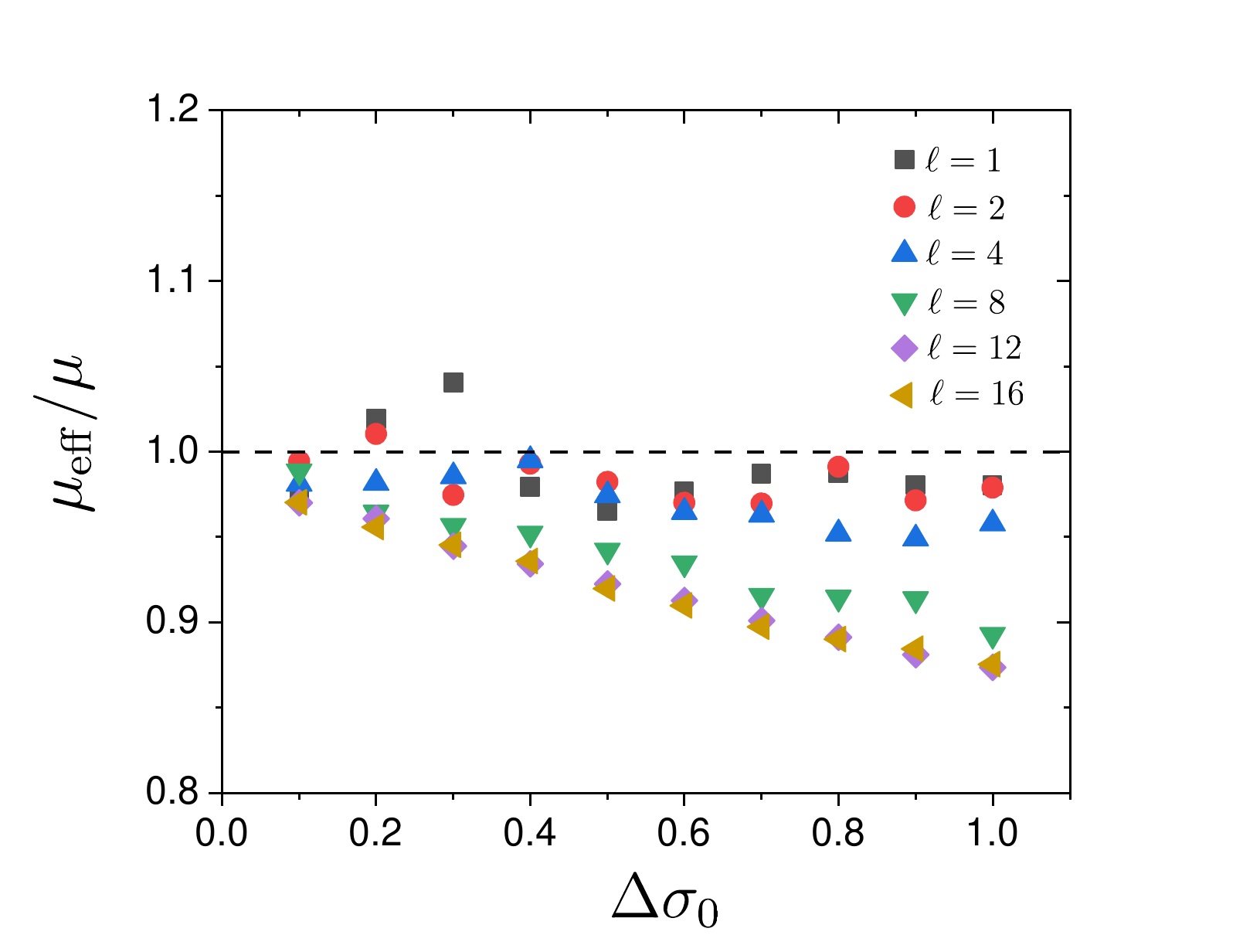}
\caption{Ratio $\mu_{\rm eff}/\mu$ versus the defect size $\ell$ for several values of perturbation amplitude $\mu\Delta\sigma_0$. The few outliers above $1$ are for small values of the defect size and small values of $\Delta\sigma_0$ where the precision of the extracted value of $\mu_{\rm eff}$ is very poor.}
\label{fig_ratio_mu}
\end{figure}

To start with, we consider if and how elasticity is renormalized due to the presence of a nonzero density of plastic events. 
Using the known expression of the 2-dimensional Eshelby propagator and considering the continuum limit, one finds that in the range $\ell/2\ll r\ll L/2$, the angle-averaged elastic component of the stress field behaves as
\begin{equation}
\Delta\sigma^{\rm elastic}(r)\approx \mu\Delta\sigma_0 \left( \frac{\ell}{2r}\right)^2\left[1+\mathcal{O}\left( \left(\frac{\ell}{2r}\right)^2\right)\right],
\label{eq_Delta_sigma_elastic_continuum}
\end{equation}
as detailed in Appendix~\ref{sec:elastic_response}.

If the presence of quadrupolar plastic events only renormalizes the elastic coupling constant, here, $\mu$, the angle-averaged total stress field should behave as 
\begin{equation}
\label{eq_Delta_sigma_full_continuum}
\overline{\Delta\sigma}(r) \approx \mu_{\rm eff} \Delta\sigma_0 \left(\frac{\ell}{2r}\right)^2 \left[1+\mathcal{O}\left(\left(\frac{\ell}{2r}\right)^2\right)\right].
\end{equation}
We expect that, in the present system which is described by elasticity only at large enough distance, there should be corrections to the above formula when $r$ is close to the boundary $\ell/2$ (and possibly when the amplitude of the perturbation $\Delta\sigma_0$ varies). There may also be finite-size effects when $r$ approaches the half-size of the system $L/2$. 

We display in Fig.~\ref{fig_renormalization} the full angle-averaged induced stress $\overline{\Delta\sigma}(r)$ and its elastic and plastic components, normalized by the amplitude of the perturbation $\mu \Delta\sigma_0$, for several values of the defect size $\ell$ and of $\mu \Delta\sigma_0$. We consider our largest system size $L=512$. To focus on the putative leading power-law dependence in the form given in Eqs.~(\ref{eq_Delta_sigma_elastic_continuum}) and (\ref{eq_Delta_sigma_full_continuum}), we use a log-log plot. On such a plot, we see that there is indeed a range of distance,  extending roughly from 2-3 to somewhat beyond $10$ times $(\ell/2)$, for which the leading behavior is proportional to $[r/(\ell/2)]^{-2}$. Throughout most of this range, the plastic component is nonzero (its maximum amplitude is around $r/(\ell/2)\approx 2$), while it reaches zero when $r/(\ell/2)\gtrsim 10$. This decay to zero is again a sign that an anomalous elasticity can only be expected over a distance of the order of the perturbation size $\ell$ and vanishes in the thermodynamic limit. 

We then extract the coefficient of the power-law dependence.  Having identified on the log-log plot the regime with a  slope of $-2$, we fit the data to obtain the amplitude of the power-law decay. Some care is required because we are using a lattice description instead of a continuum one. As a result, both Eqs.~(\ref{eq_Delta_sigma_elastic_continuum}) and (\ref{eq_Delta_sigma_full_continuum}) are corrected by the same overall factor that accounts for the modified discrete Eshelby kernel (see Appendix~\ref{sec:elastic_response}) and the fact that the central defect is not quite circular when $\ell$ is small. However, this overall factor cancels in the ratio $\mu_{\rm eff}/\mu$, which is represented as a function of $\Delta\sigma_0$ for different values of $\ell$ in Fig.~\ref{fig_ratio_mu}. This figure shows that in the range over which plastic activity induced by the initial Eshelby perturbation takes place, a range that scales with $\ell$, there is indeed a renormalization of the effective elastic constant. Yet, this renormalization depends on the size and the amplitude of the perturbation.

\begin{figure}
\includegraphics[width=.8\linewidth]{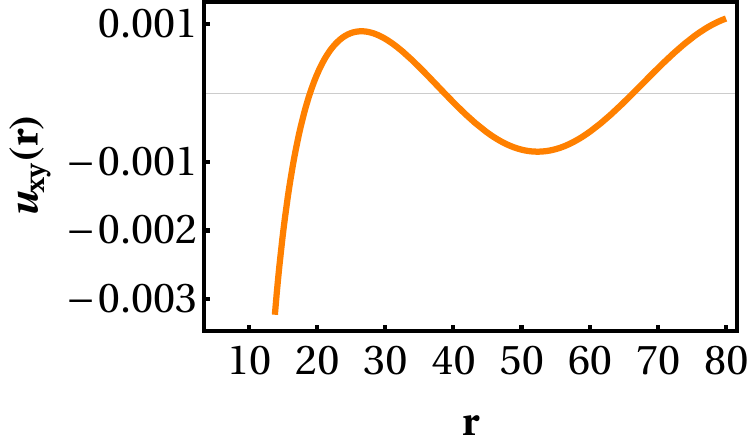}
\caption{Atomic simulation: Radial component of the XY strain field as a function of the distance $r$ for the Eshelby problem with $r_{\rm in}=5$, $r_{\rm out}=80$, and $\delta=1.05$ (see Fig.~\ref{fig_sketch-Eshelby}(a)). The XY strain field is obtained by combining simulation results for the displacement field in a poorly annealed glass and an analytical description of anomalous 
elasticity.\cite{hentschel_eshelby,avanish-eshelby} 
}
\label{fig_radial-strain-field_avanish}
\end{figure}

\begin{figure}
\includegraphics[width=.87\linewidth]{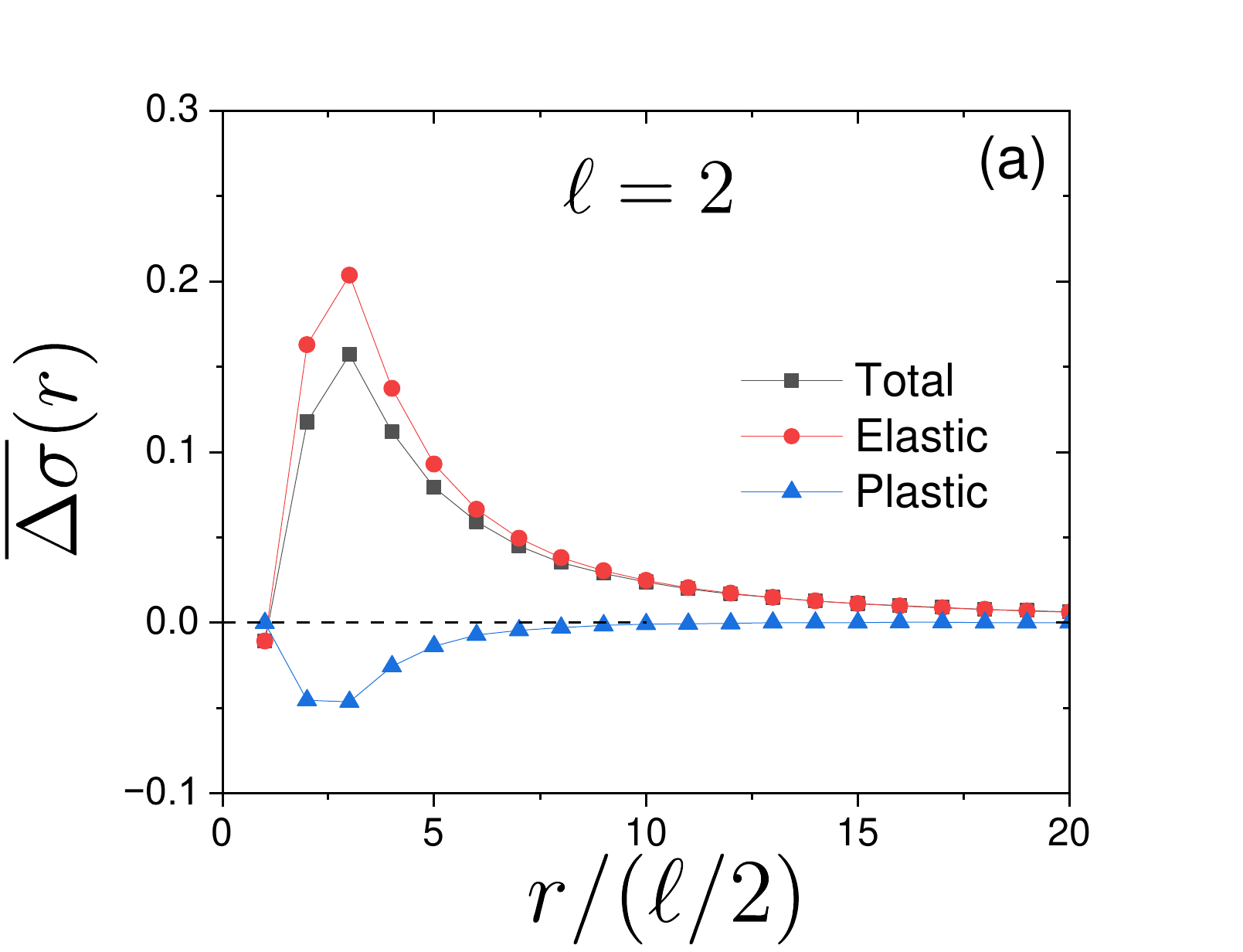}
\includegraphics[width=.87\linewidth]{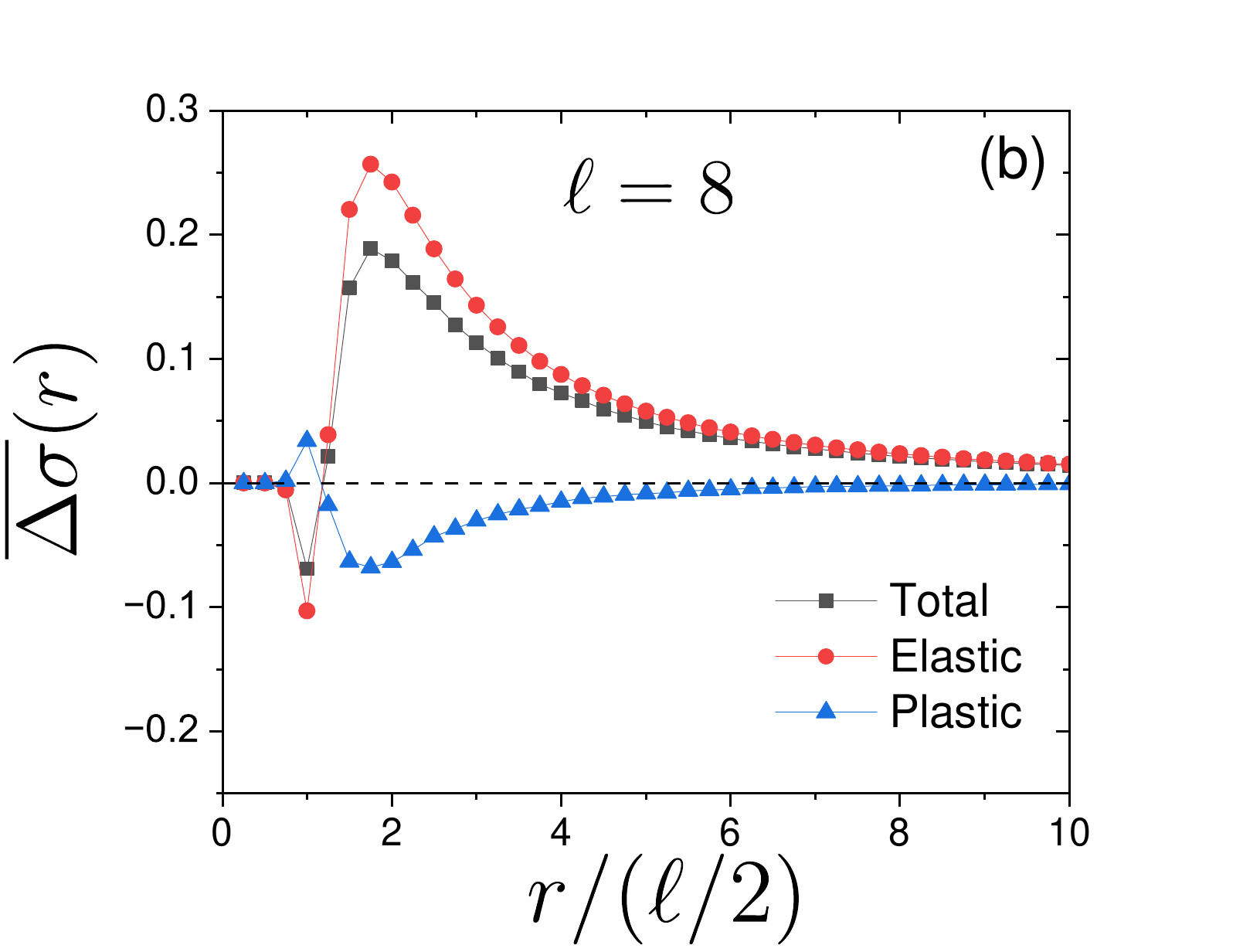}
\includegraphics[width=.87\linewidth]{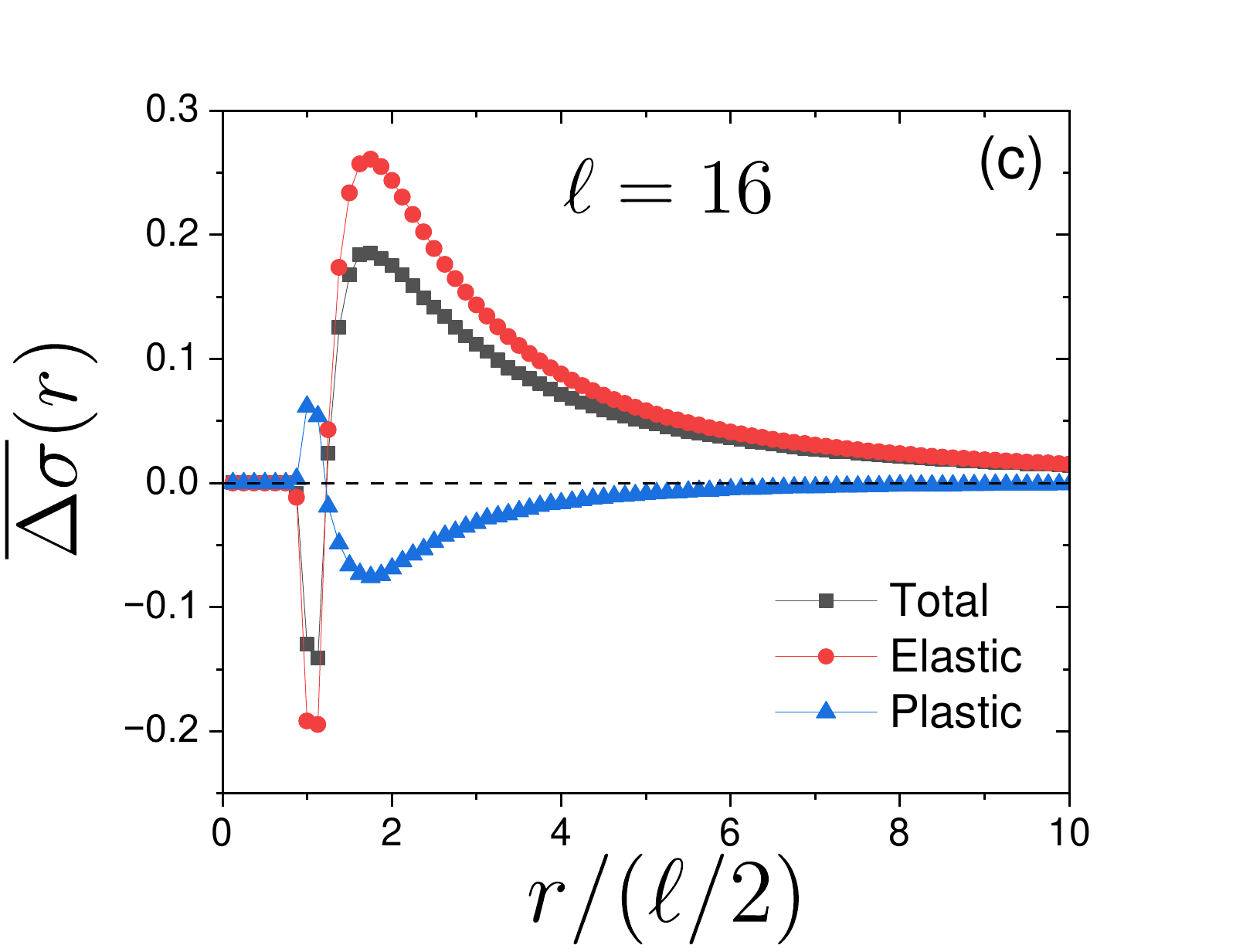}
\caption{Elasto-plastic model: Radial component of the total XY stress field $\overline{\Delta\sigma}(r)$ and its plastic and elastic components, $\overline{\Delta\sigma}^{\rm plastic}(r)$ and $\overline{\Delta\sigma}^{\rm elastic}(r)$, as a function of the normalized distance $r/(\ell/2)$ for the Eshelby problem with 
$L=512$ and three values of the initial defect size, $\ell=2$ (top), $\ell=8$ (middle), and $\ell=16$ (bottom). 
}
\label{fig_dipole-screening_EPM}
\end{figure}

\begin{figure}
\includegraphics[width=.87\linewidth]{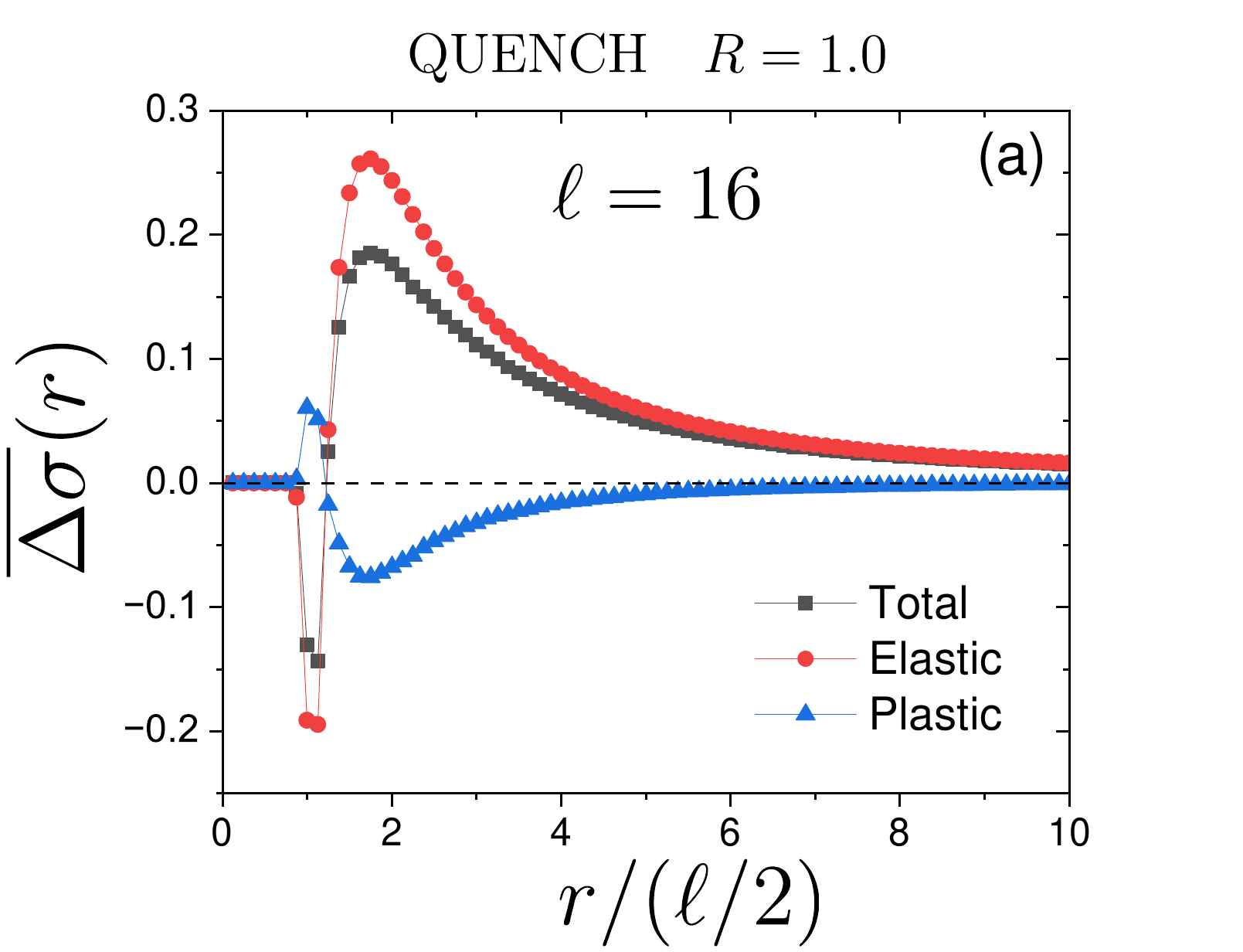}
\includegraphics[width=.87\linewidth]{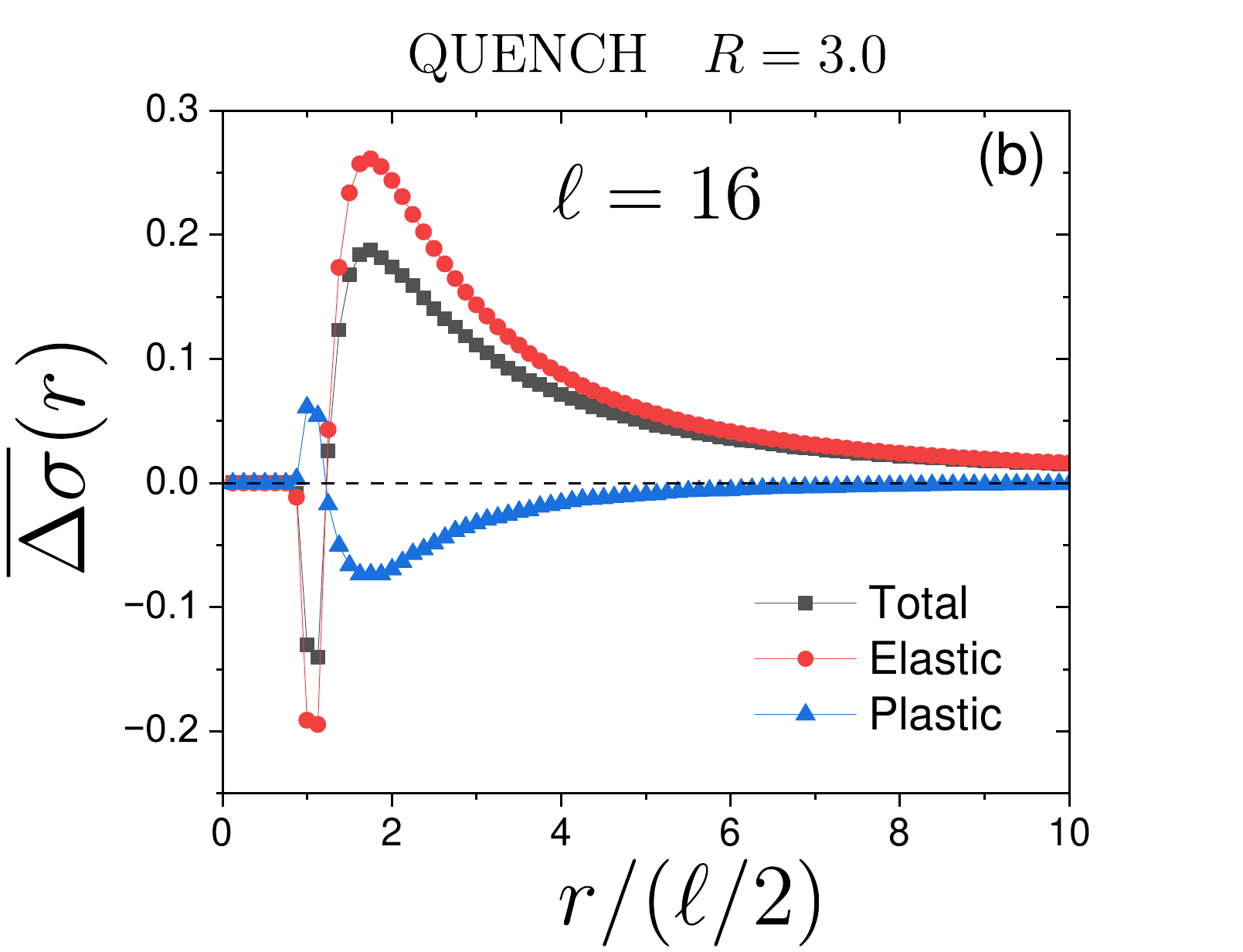}
\includegraphics[width=.87\linewidth]{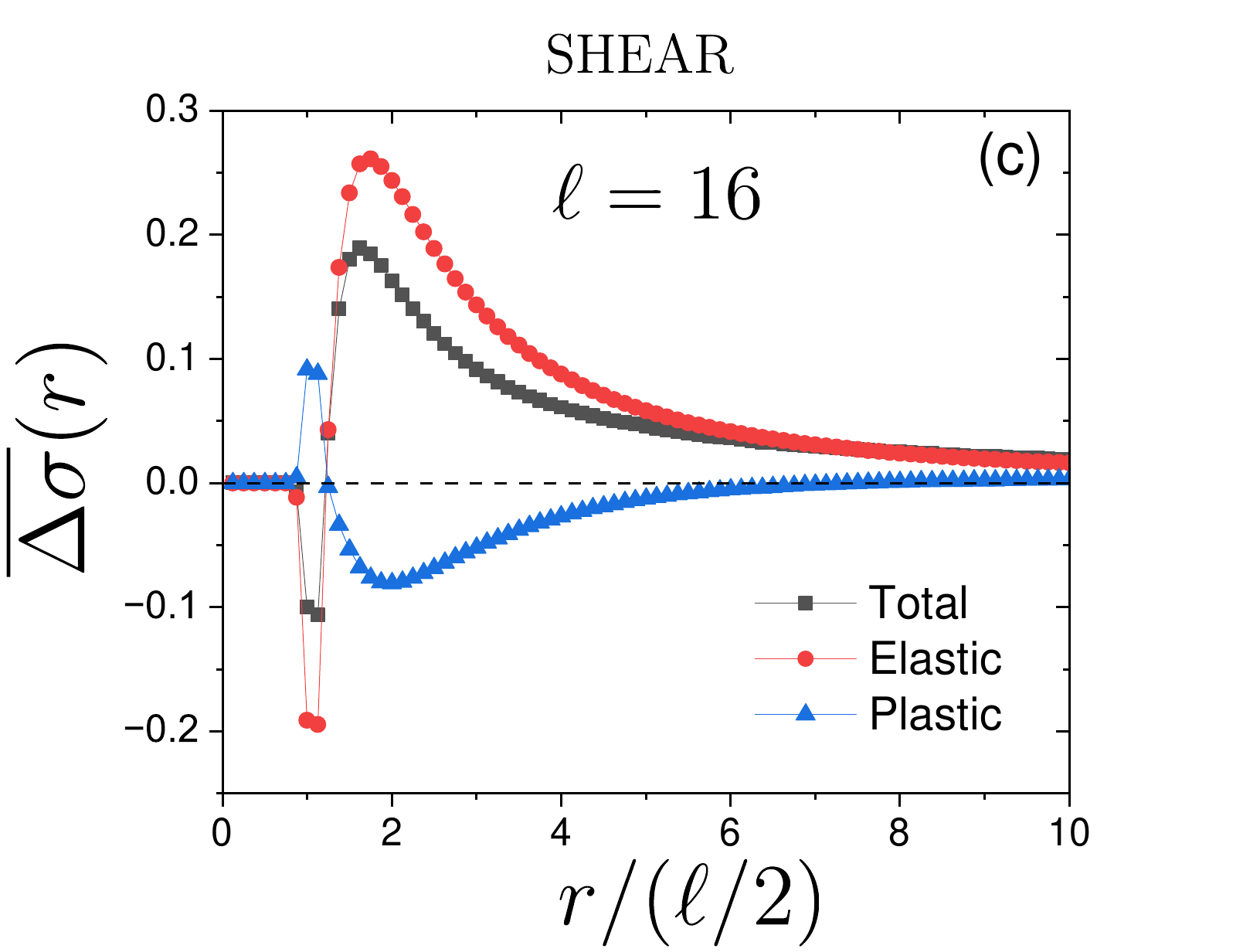}
\caption{Elasto-plastic model: Radial component of the total XY stress field $\overline{\Delta\sigma}(r)$ and its plastic and elastic components, $\overline{\Delta\sigma}^{\rm plastic}(r)$ and $\overline{\Delta\sigma}^{\rm elastic}(r)$, as a function of the normalized distance $r/(\ell/2)$ for the Eshelby problem with 
$L=256$ and $\ell=16$ for three different initial preparation protocols.
}
\label{fig_different_protocols}
\end{figure}

Finally, we turn to the potential screening by the dipole field formed by the divergence of the quadrupole field. As we have already mentioned, elasto-plastic models do not describe the displacement field, and the consequences of dipole screening should rather be checked in the (XY) stress field (or strain field if the two are assumed to be proportional). 

The outcome for the radial component of the XY strain field $u_{\rm XY}(r)$ obtained from an atomic simulation of the Eshelby problem combined with an analytical description of the anomalous elasticity\cite{hentschel_eshelby,avanish-eshelby} is shown in Fig.~\ref{fig_radial-strain-field_avanish}. The signature of dipole screening is the nonmonotonic dependence on distance, with a strain that changes sign. A purely elastic response would indeed lead to a monotonic dependence and no change of sign. Note that the amplitude of the modulation is very small and that the boundary condition at the outer circle for $r=r_{\rm out}$ is such that the displacement field vanishes, the strain then being nonzero.

We show the results of our elasto-plastic model in Fig.~\ref{fig_dipole-screening_EPM} for three defect sizes $\ell=2,8$, and $16$. We display the data on a lin-lin plot to focus on the region where plastic activity is nonnegligible (compare with the right panels of Fig.~\ref{fig_renormalization}). As expected, beyond the first layer close to the initial central defect, {\it i.e.}, for $r\geq \ell$, the elastic component of the stress field always decays monotonically and does not change sign. (Note the difference of symmetry axes and signs between the results of our elasto-plastic model and those of the simulation of [\onlinecite{hentschel_eshelby,avanish-eshelby}]; this is, however, immaterial to the discussion.) 

What we also find is that, in the same range, the total stress field and its plastic component also appear to show monotonic behavior with no change in sign. This seems to be true regardless of the size of the initial Eshelby perturbation. As one can see from Fig.~\ref{fig_different_protocols}, this is found as well for all protocols that we have followed to prepare the initial configurations.

The above results indicate that, while elasto-plastic models can capture anomalous elasticity in the form of quadrupole screening leading to renormalized elasticity, they do not seem to capture dipole screening. This absence of dipole screening may be due to several factors:

- The amorphous solids prepared in atomic simulations seem to cover a wider range of disorder strength, stability, and proximity to the jamming transition (if present) than the systems that are used as initial conditions in elasto-plastic models (whatever the preparation protocol); as a result, the density of plastic events generated in the latter is less than what can be achieved in atomic simulations.

- The boundary conditions used in [\onlinecite{hentschel_eshelby,avanish-eshelby}] are different than the periodic boundary conditions of the elasto-plastic model; this could explain the difference in the number of modulations seen in the radial component of the strain or stress field because the strain and stress go to 0 at large distance in the EPM while they go to nonzero values in the simulation.

- Stress, as studied in the elasto-plastic model, and strain, as derived in the simulation and theory, may actually not be straightforwardly compared when there is dipole screening. In the case of quadrupole screening, they are linearly related by the renormalized elastic constants; but what in the case of dipole screening?

- More importantly, we have argued that many of the ingredients for anomalous elasticity are present in elasto-plastic models, in particular quadrupole and dipole fields associated with the plastic events. However, it could be that the feedback effect of the plastic events onto the elastic behavior is not properly encoded in the kinetic rules of the elasto-plastic models.
\\

\section{Conclusion}
\label{sec_conclusion}

In this work, we have studied anomalous elasticity in amorphous solids, focusing on the conditions under which a finite density of quadrupolar Eshelby-like singularities emerges under mechanical loading, and the extent to which this leads to deviations from classical elasticity theory. By combining general theoretical arguments and simulations using an elasto-plastic model in the athermal quasi-static limit, we have clarified to what extent anomalous elasticity manifests itself as a true breakdown of long-wavelength elasticity. In particular, we have clarified the role of finite-size effects, and shown that the extent over which one finds a density of quadrupoles generically scales as the size of the mechanical perturbation. Our analysis has also examined whether elasto-plastic models can qualitatively capture the key features of anomalous elasticity, including elastic constant renormalization and dipole screening effects, and how these compare with predictions from atomistic simulations and continuum approaches. 

We conclude by highlighting two directions for future research that naturally follow from our results. The first one concerns the inability of elasto-plastic models to capture anomalous elasticity in the form of dipole screening, if present. We have outlined several possible reasons for this failure, the most important being that the elastic field is not consistently treated in these models. In particular, there is no proper feedback between plastic relaxation and elastic deformations. If confirmed, this would highlight a clear weakness of elasto-plastic models and, consequently, it would point towards an important extension needed for these models. From this perspective, it would also be interesting to check whether the more refined versions of elasto-plastic models already present in the literature [\onlinecite{EPM_review}] can capture a potential dipole screening effect.

The second point worth investigating further concerns the long-wavelength theory associated with anomalous elasticity. Once it is established that classical long-wavelength elasticity breaks down in amorphous materials, the natural question is: what theoretical framework replaces it? A significant step was made by Procaccia and coworkers \cite{lemaitre_anomalous,anomalous_review}, who demonstrated how classical elasticity fails in the presence of a finite (nonzero)  density of quadrupolar plastic events. However, in their approach, the quadrupolar density field is introduced as an external input. A complete anomalous long-wavelength theory would ideally describe the mutual feedback between the elastic fields and the quadrupolar density field.  As we have shown, a mechanical perturbation of size $\ell$ generates a surrounding region of quadrupolar activity of the same scale, which in turn modifies the elastic response. This suggests that a full theory should combine the framework developed by Procaccia and coworkers with an additional component that predicts the induced quadrupolar density field for a given perturbation.\textcolor{red}{\cite{footnote_nonlinear}} Obtaining such a full theory is a very interesting direction for future work.  

\acknowledgements{We thank Avanish Kumar and Itamar Procaccia for many fruitful exchanges and for providing us with analytical and simulation data on the Eshelby problem. We also thank Bulbul Chakraborty for interesting discussions.}

\appendix

\section{The elasto-plastic model and the setup}
\label{app_EPM}

\subsection{Model}
\label{sec:model}

We study a scalar elasto-plastic model~\cite{rossi_EPM} in a two-dimensional lattice with periodic boundary conditions. The linear box length is $L$, using the lattice constant as the unit of length. For each site ${\bf r}_i$ (corresponding to the center of a "block"), we assign a local shear stress $\sigma_i$ (scalar variable) and a plastic activity indicator $n_i=0,1$.

If $\vert \sigma_i\vert $ is greater than or equal to a threshold $\sigma^{\rm th}>0$, $|\sigma_i| \geq \sigma^{\rm th}$, this site undergoes a plastic event: $n_i=0\to n_i=1$ and $\sigma_i \to \sigma_i - \Delta \sigma_i$, where $\Delta \sigma_i$ is the local stress drop associated with the plastic rearrangement. We use a uniform threshold that defines the unit of stress, $\sigma^{\rm th}=1$. The stress drop $\Delta \sigma_i$ is a stochastic variable. In this study, we use $\Delta \sigma_i=(z+|\sigma_i|-\sigma^{\rm th}){\rm sgn}(\sigma_i)$, where ${\rm sgn}(x)$ is the sign function and $z >0$ is a random number drawn from an exponential distribution, $p(z)=\frac{1}{z_0} e^{-z/z_0}$. The mean value $z_0$ is set to $z_0=1$. 

A local plastic event at site $i$ influences the local stress at all other sites ($\forall j \neq i$) according to 
\begin{equation}
    \sigma_j \to \sigma_j + {G}^{\rm Eshelby}_{\psi_i}({\bf r}_{ji}) \ \Delta \sigma_i,
\end{equation}
where ${\bf r}_{ji}={\bf r}_{j}-{\bf r}_{i}$ and $\psi_i$ is the orientation of the Eshelby kernel with respect to the horizontal $x$-axis. 
In the problem studied here, we consider that all local Eshelby quadrupoles are aligned with the original perturbation (unless otherwise stated, see below), with $\psi_i=0$ $\forall i$ such that $n_i=1$. The numerical implementation of ${G}^{\rm Eshelby}_{\psi_i}({\bf r})$ is described in [\onlinecite{rossi_EPM}].

\subsection{Initial conditions}

We prepare initial glass samples according to two different classes of protocols before applying a perturbation at the center of the system.

The first preparation method generates an isotropically quenched configuration, obtained from a Gaussian distribution of local stresses. We call this protocol QUENCH. We start by assigning local stresses drawn from a Gaussian distribution  with zero mean and standard deviation $R$. As a result, some sites may have stresses exceeding the threshold, {\it i.e.}, $|\sigma_i| > \sigma^{\rm th}$. From this initial state, we randomly select a site $i$ among the $L^2$ sites and apply the kinetic rules described in Sec.~\ref{sec:model}, which include local stress drops and stress redistribution via the Eshelby kernel. During the quenching process, we also randomly assign the angle $\psi_i \in [0, \pi/2)$ for each plastic activity to mimic a glass sample obtained by quenching from an isotropic liquid. We repeat this Monte-Carlo-like update until all sites are stable, which means $|\sigma_i| \leq \sigma^{\rm th}$. This final stable configuration is then used as the initial glass sample for further perturbation.

To assess the effect of the disorder, we study glass samples generated using Gaussian distributions with $R = 1.0$ and $R = 3.0$. For $R = 1.0$, we generated 1000 independent initial configurations for system sizes $L = 64$ and $128$, and 100 configurations for $L=256$ and 512. For $R = 3.0$, we generated 100 configurations for $L = 256$. In the main text, we present results mainly for $R = 1.0$, unless otherwise stated.

We also consider initial glass samples obtained from steady-state shearing. We call this protocol SHEAR. In particular, we perform strain-controlled simulations~\cite{rossi_EPM} (with the angle fixed as $\psi_i = 0$ for all $i$), starting from a Gaussian distribution with $R = 1.0$. We confirm that the system reaches a steady state when the applied strain $\gamma$ becomes large enough, $\gamma = 5.0$. The final configuration at $\gamma = 5.0$ is used as the initial glass sample for further analysis.
We generated 100 independent initial configurations for $L = 256$.
Note that this glass sample is no longer isotropic, as seen from the asymmetry in the local stress distribution inclining toward the positive axis. Nevertheless, we find that the results obtained for the sheared initial glass samples are qualitatively similar to those for the isotropic quenched one (see Fig.~\ref{fig_different_protocols}).

\subsection{The setup for the Eshelby problem}

Once an initial glass configuration is obtained, either through the QUENCH or the SHEAR protocol, we define a circular defect region $\mathcal{D}$ in the center of the sample, as shown in Fig.~\ref{fig_sketch-Eshelby}(b). The diameter of the defect is denoted by $\ell$. (Note that for a small $\ell$ the deviation from a circle due to the underlying lattice is significant.)

We then artificially induce plastic events within the defect region $\mathcal{D}$ by applying a uniform stress drop of magnitude $\mu \Delta \sigma_0$. We vary the magnitude of the stress drop in the range $\mu \Delta \sigma_0 = 0.1$--$1.0$.
 The total stress redistribution is obtained by adding contributions from all sites within the defect. As a result of the initial stress drop in the central domain, stress redistribution occurs due to the Eshelby kernel with a fixed orientation.

This perturbation induces instabilities (plastic events) at various sites outside the defect. These secondary plastic events can, in turn, trigger additional events, forming an avalanche.
For these subsequent plastic events, the stress drops $\Delta \sigma_i$ are randomly assigned, while the orientation of the Eshelby kernel remains fixed. We note that after the initial perturbation corresponding to the stress drop 
$\mu \Delta \sigma_0$, the defect region is frozen and does not exhibit any subsequent plastic event. This assumption is consistent with experimental setups and molecular dynamics simulations reported by Procaccia and co-workers.

The results exhibit strong sample-to-sample fluctuations. 
To account for this variability, we performed disorder averaging over many independent samples. In practice, starting from a given initial configuration, we randomly selected ten different locations for the center of the defect. This procedure yielded $10^4$ samples for $L = 64$, $128$ and $10^3$ samples for $L=256$,  $512$ in the QUENCH protocol with $R = 1.0$, $10^3$ samples for $L = 256$ in the QUENCH protocol with $R = 3.0$, and for $L = 256$ in the SHEAR protocol.

\section{Additional technical details}

\subsection{Numerical computation of observables}
\label{sec:numerical_computation}

Here, we explain how to compute $\overline N(r)$ defined in Eq.~(\ref{eq:bar_N}) and $\overline{\Delta\sigma}(r)$ defined in Eq.~(\ref{eq:Delta_sigma_average}) for a two-dimensional lattice.

First, in the  two-dimensional polar coordinates, ${\bf r}=(r,\theta)$, the delta function is expressed as
\begin{equation}
\delta^{(2)}({\bf r}-{\bf r}_i)=r^{-1}\delta(r-r_i)\delta(\theta-\theta_i) . 
\label{eq:delta_function_transform}
\end{equation}
In consequence, with Eqs.~(\ref{eq:n_r}) and (\ref{eq:bar_N}), $\overline N(r)$ is given by 
\begin{equation}
    \overline N(r) = \sum_{i\notin \mathcal D} \overline n_i \ \chi_{[\ell/2, \ r]}(r_i) ,
    \label{eq:bar_N_numerical}
\end{equation}
where $\chi_{[a, \ b]}(x)$ is an indicator function defined as $\chi_{[a, \ b]}(x)=1$ if $a \leq x \leq b$, $\chi_{[a, \ b]}(x)=0$ otherwise.
Equation~(\ref{eq:bar_N_numerical}) is used in numerical simulations.
\\

On the other hand, $\overline{\Delta\sigma}(r)$ in Eq.~(\ref{eq:Delta_sigma_average}) is given by
\begin{equation}
    \overline{\Delta\sigma}(r)=\sum_{i \notin \mathcal D} \overline{\Delta\sigma_i}^{\rm f/i} \frac{\cos(4 \theta_i)}{\pi r_i} \delta(r-r_i) .
    \label{eq:Delta_sigma_platic_numerical}
\end{equation}
This equation is numerically evaluated by a kernel density estimation,
\begin{equation}
    \delta(r-r_i) \simeq \frac{1}{\sqrt{2 \pi \epsilon^2}} e^{-\frac{(r-r_i)^2}{2 \epsilon^2}} ,
\end{equation}
where $\epsilon$ is a smoothing parameter, which is set to $\epsilon=0.5$.
We checked that a range of choices, $\epsilon=0.3-1.0$, does not change the result.

$\Delta\sigma^{\rm elastic}(r)$ is computed in the same way as $\overline{\Delta\sigma}(r)$. We then obtain $\overline{\Delta\sigma}^{\rm plastic}(r)$ as a difference, $\overline{\Delta\sigma}^{\rm plastic}(r)=\overline{\Delta\sigma}(r)-\Delta\sigma^{\rm elastic}(r)$.

\subsection{Multiple local yielding}
\label{sec:multiple_yielding}

\begin{figure}
\includegraphics[width=.9\linewidth]{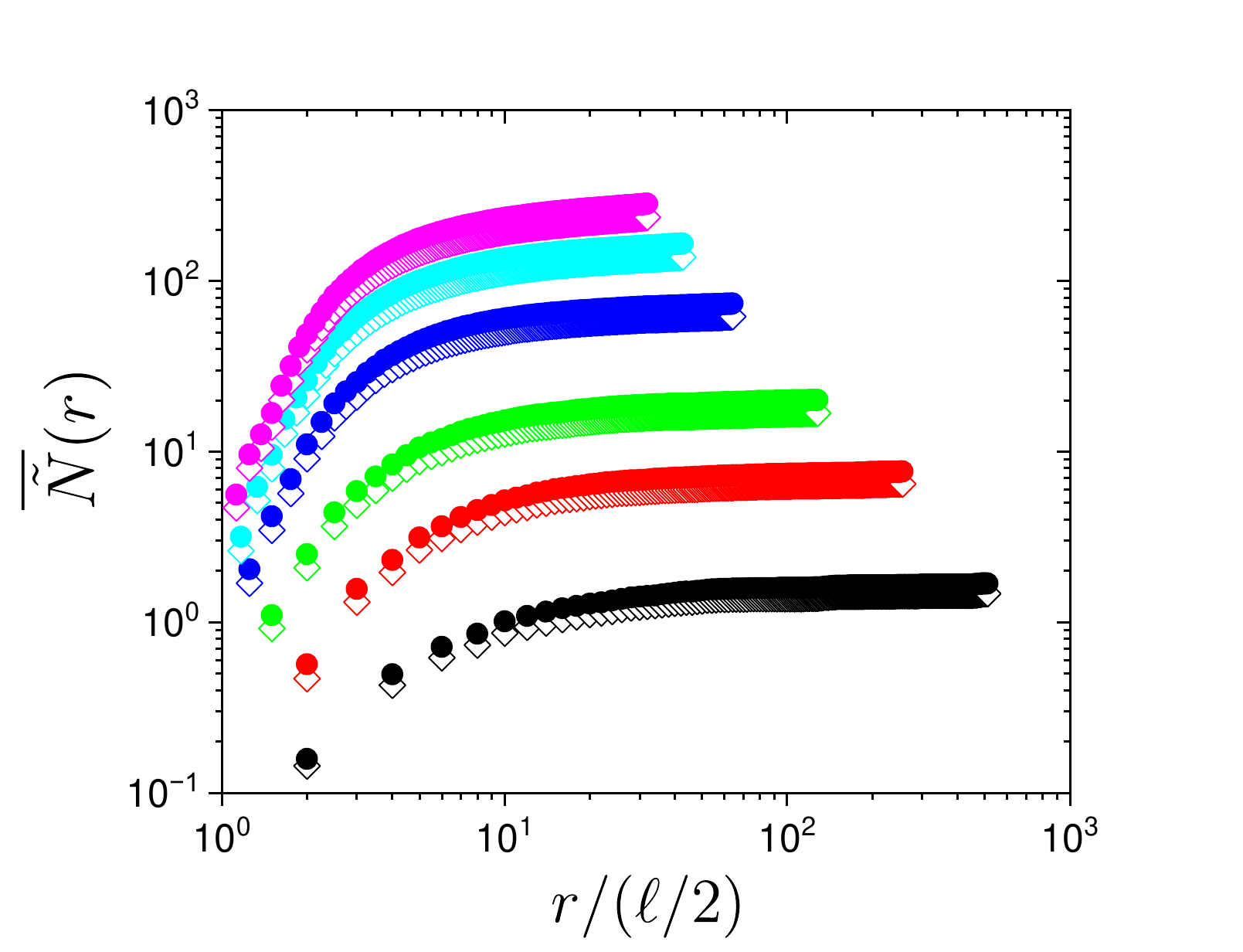}
\caption{Comparison between the averaged number of sites $\overline{N}(r)$ that have yielded at least once during the evolution after the perturbation (empty diamonds) and the averaged number of plastic events including multiple yielding sites $\overline{\widetilde N}(r)$ (filled circles) versus $r/(\ell/2)$  for several values of the initial defect size: from bottom to top, $\ell=1$ (black), 2 (red), 4 (green), 8 (blue), 12 (light blue), and 16 (pink). The system size of $L=512$.
}
\label{fig:binary_or_accumulation}
\end{figure}

As alluded in the main text, in an elasto-plastic model local yielding events may appear and relax at each step of the algorithm in the evolution between the initial and the final states. As a result, a given site can in principle yield, {\it i.e.}, be plastically active, several times. To assess the effect of this multiple local yielding, we have monitored the accumulated plastic activity $\widetilde n_i$ at each site, where $\widetilde n_i$ counts the number of times the site $i$ has yielded. We compute from it $\overline{\widetilde N}(r)$, the sample-averaged number of plastic events (including now multiple yielding) in a circular domain of radius $r$ centered on the origin, but excluding the initial Eshelby defect $\mathcal D$. In Fig.~\ref{fig:binary_or_accumulation} we compare the outcome with the sample-averaged number of plastically active sites $\overline{N}(r)$ (without taking into account multiple yielding). One can see that the effect of multiple yielding at a given site is small and does not change the overall behavior.

\section{Elastic response from a circular defect}
\label{sec:elastic_response}

We start with the continuum description of the elastic response from a point defect studied by Picard et al.,\cite{picard_scalar} where the defect is modeled by a delta function,
\begin{equation}
    \Delta \sigma^{\rm elastic}({\bf r}) = \int d {\bf r}' G^{\rm Eshelby}({\bf r}-{\bf r}')\mu\Delta \sigma_0 \delta^{(d)}({\bf r}') ,
\end{equation}
where $\Delta \sigma_0$ is the magnitude of the stress drop and $\int d {\bf r}'\equiv \int d^d r'$ with $d$ the dimension of space. We extend this to the case of finite-size defects, in particular, to defects having a spherical shape,
\begin{equation}
    \Delta \sigma^{\rm elastic}({\bf r}) = \int d {\bf r}' G^{\rm Eshelby}({\bf r}-{\bf r}')\mu\Delta \sigma_0 I({\bf r}') ,
\end{equation}
where $I({\bf r})=\theta(\ell/2-r)$ is an indicator function with the radius of the defect $\ell/2$.

Using the Fourier transform, $\hat f({\bf q})=\mathcal{F}[f({\bf r})]=\int d {\bf r} e^{-i{\bf q}\cdot {\bf r}}f({\bf r})$, we obtain the following expression,
\begin{equation}
    \Delta \widehat \sigma^{\rm elastic}({\bf q}) = \widehat G^{\rm Eshelby}({\bf q}) \mu\Delta \sigma_0 \widehat I({\bf q}) .
\end{equation}
In $d=2$, ${\bf q}=(q_x, q_y)$, $\widehat G^{\rm Eshelby}({\bf q})$, and $\widehat I({\bf q})$ are given by
\begin{eqnarray}
    \widehat G^{\rm Eshelby}({\bf q}) &=& - \frac{4 q_x^2 q_y^2}{q^4} , \\
    \widehat I({\bf q}) &=& \frac{\pi \ell}{q} J_1(q \ell/2) ,    
\end{eqnarray}
where $J_1(x)$ is the Bessel function of the first kind of order 1. 

When $r>\ell/2$, we arrive at
\begin{eqnarray}
    \Delta \sigma^{\rm elastic}({\bf r}) &=& \mathcal{F}^{-1}\left[\widehat G^{\rm Eshelby}({\bf q})\mu \Delta \sigma_0 \widehat I({\bf q})\right] \nonumber \\
    &=& \mu\Delta \sigma_0 \cos(4\theta)\left[ \left( \frac{\ell}{2r}\right)^2 - \frac{3}{2}\left( \frac{\ell}{2r}\right)^4 \right] . \nonumber \\
\end{eqnarray}

The radial component obtained by averaging over the polar angle is then given by
\begin{eqnarray}
    \Delta\sigma^{\rm elastic}(r) &=& \frac 1{\pi}\int_0^{2\pi} d\theta \cos(4\theta) \Delta\sigma^{\rm elastic}({\bf r}) \nonumber \\
    &=& \mu\Delta \sigma_0 \left[ \left( \frac{\ell}{2r}\right)^2 - \frac{3}{2}\left( \frac{\ell}{2r}\right)^4 \right] .
    \label{eq:Delta_sigma_elastic_solution}
\end{eqnarray}

\begin{figure}
\includegraphics[width=.9\linewidth]{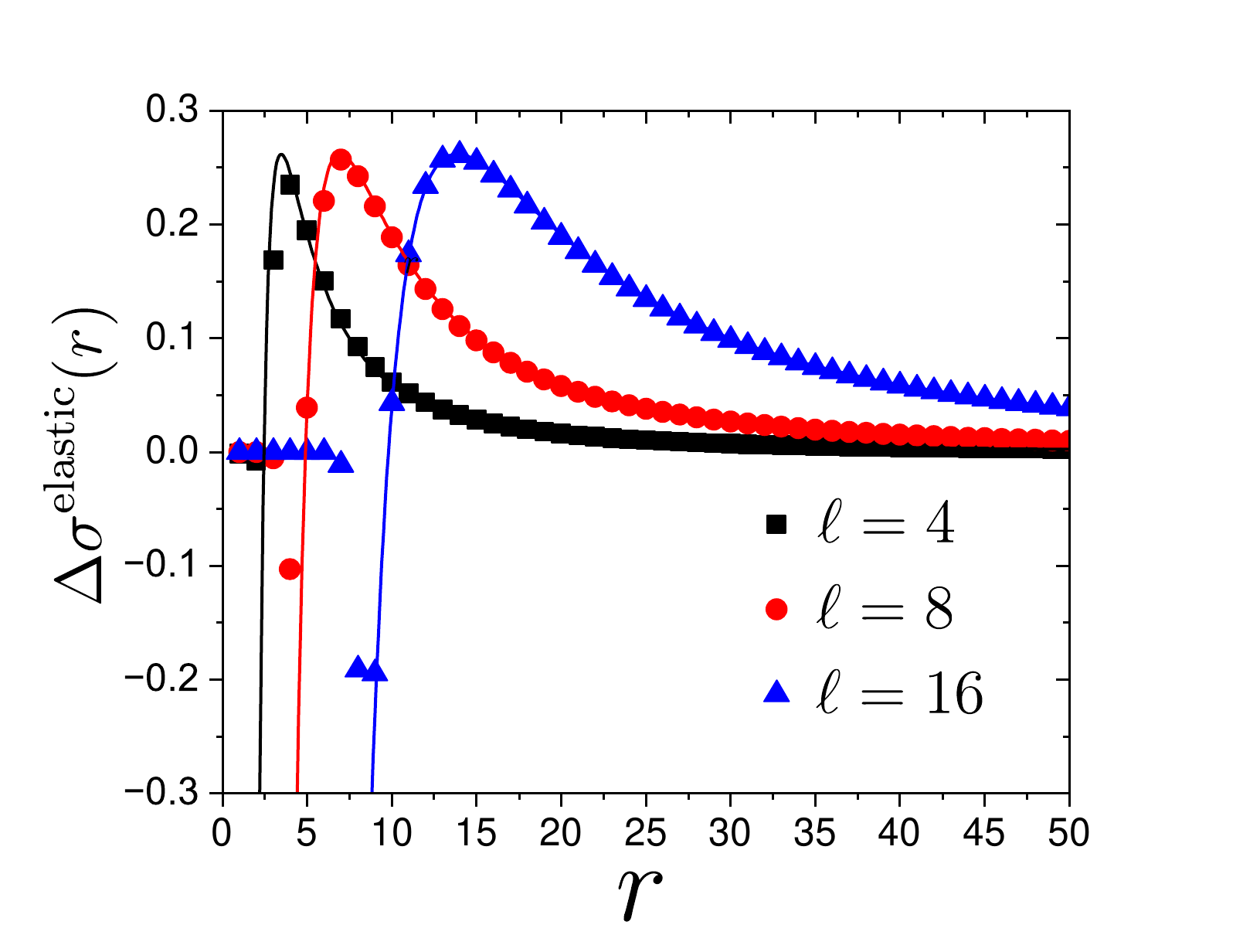}
\caption{Comparison between $\Delta \sigma^{\rm elastic}(r)$ obtained by numerical simulations on the lattice (points) and the continuum linear-elasticity solution (solid curves) in Eq.~(\ref{eq:Delta_sigma_elastic_solution}). The system size is $L=256$. To perform the comparison properly, the continuum solutions are divided by the normalization factor $\mathcal{G} \simeq 0.64$ for the discrete Eshelby kernel~\cite{rossi_EPM}.}
\label{fig:Delta_sigma_elastic}
\end{figure}

Figure~\ref{fig:Delta_sigma_elastic} compares $\Delta \sigma^{\rm elastic}(r)$ obtained by numerical simulation on the discrete lattice and the continuum linear-elasticity solution in Eq.~(\ref{eq:Delta_sigma_elastic_solution}) for $\mu\Delta \sigma_0=1$. (Note that there is an overall normalization factor when going from the continuum to the discrete Eshelby kernel~\cite{rossi_EPM}.) Near the circular defect, $r \simeq \ell/2$, a negative part appears in numerical simulations, and its magnitude increases with the defect size. This is in agreement with the continuum solution.

\begin{thebibliography}{15}
\bibitem{footnote_anomalous}
Anomalous linear elasticity is also found in disordered networks close to their rigidity transition,\cite{lerner-bouchbinder_23} but this corresponds to a different phenomenon that we do not discuss here.
\bibitem{footnote_nonlinear}
There is a recent nonlinear extension of the theory of Kumar and Procaccia\cite{procaccia_nonlinear} to describe the appearance of shear banding during a minimization step in an AQS shear protocol. However, this does not predict the kinetic evolution among many different mechanical minima encountered during the whole shear experiment.
\end{thebibliography}
\end{document}